\documentclass{article}
\usepackage[utf8]{inputenc}
\usepackage{amsmath}
\usepackage{amsthm}
\usepackage{amsfonts}
\usepackage{algorithm}
\usepackage{algorithmic}
\usepackage{graphicx}
\usepackage[colorinlistoftodos]{todonotes}
\usepackage{fullpage}
\usepackage{bbm}
\usepackage{subcaption}
\usepackage{tikz}
\usepackage[normalem]{ulem}
\usepackage{nicefrac}

\newtheorem{theorem}{Theorem}
\newtheorem{example}[theorem]{Example}

\newtheorem{definition}[theorem]{Definition}
\newtheorem{proposition}[theorem]{Proposition}

\newtheorem{lemma}[theorem]{Lemma}
\newtheorem{corollary}[theorem]{Corollary}

\newtheorem{thmx}{Theorem}
\newtheorem{propx}[thmx]{Proposition}

\newcommand{\R}{\mathbb{R}}
\newcommand{\E}{\mathbb{E}}
\renewcommand{\P}{\mathbb{P}}

\renewcommand{\S}{\mathcal{S}}

\newcommand{\Obj}{\textsf{Obj}}
\newcommand{\Alg}{\textsf{Alg}}

\renewcommand{\c}{\hat{c}}

\allowdisplaybreaks

\title{Secretaries with Advice}
\author{Paul D\"utting\thanks{Google Research Z\"urich, Switzerland, \texttt{duetting@google.com}} \and Silvio Lattanzi\thanks{Google Research Z\"urich, Switzerland, \texttt{silviol@google.com}} \and Renato Paes Leme\thanks{Google Research New York, USA, \texttt{renatoppl@google.com}} \and Sergei Vassilvitskii\thanks{Google Research New York, USA, \texttt{sergeiv@google.com}}}

\date{}
\begin{document}

\maketitle
\thispagestyle{empty}

\begin{abstract}
The secretary problem is probably the purest model of decision making under uncertainty. In this paper we ask which advice can we give the algorithm to improve its success probability?

We propose a general model that unifies a broad range of problems: from the classic secretary problem with no advice, to the variant where the quality of a secretary is drawn from a known distribution and the algorithm learns each candidate's quality on arrival, to more modern versions of advice in the form of samples, to an ML-inspired model where a classifier gives us noisy signal about whether or not the current secretary is the best on the market.

Our main technique is a 
factor revealing LP that captures all of the problems above. We use this LP formulation to gain structural insight into the optimal policy. Using tools from linear programming, we present a tight analysis of optimal algorithms for secretaries with samples, optimal algorithms when secretaries' qualities are drawn from a known distribution, and a new noisy binary advice model.
\end{abstract}

\clearpage
\pagenumbering{arabic} 

%% SECTION 1
\section{Introduction}

The secretary problem captures one of the purest forms of decision making under uncertainty:  $n$ candidates arrive one at a time, and must either be accepted or rejected on the spot. The goal is to design a stopping policy that maximizes the probability of selecting the best candidate.  

Some additional information is necessary to find policies that have provable performance guarantees. A standard assumption is that candidates arrive in random order, in which case  one can achieve a $\nicefrac{1}{e} \approx 0.36$ competitive ratio by hiring the first candidate who is better than the first $\nicefrac{n}{e}$ applicants. 

However, pure random arrival is not the only plausible paradigm, and previous work considered other information augmentation schemes.
For instance, Gilbert and Mosteller~\cite{gilbert2006recognizing} posit that each candidate has a quality in $[0,1]$, which is drawn independently from some fixed and known distribution. 
Gilbert and Mosteller claimed the optimality of a certain sequence of decreasing thresholds, and showed numerically that this leads to a winning probability of $\approx 0.58$ \cite{gilbert2006recognizing}. Follow-up work formally showed that this policy is optimal and that the winning probability of the optimal policy converges to $e^{-c} + (e^c-c-1) \int_{1}^{\infty} x^{-1} e^{-cx} \;dx \approx 0.580164$ where $c\approx 0.804352$ is the unique solution to $\sum_{j = 1}^{\infty} c^j/(j!j) = 1$ \cite{BerezovskiyG84,gnedin96,Samuelson82}. 

Another natural form of advice are samples. Indeed there has been a flurry of recent work on stopping problems, especially the prophet inequality variety, with limited information about an underlying distribution in the form of samples (e.g., \cite{DBLP:conf/soda/AzarKW14,CorreaEtAl21,DBLP:conf/soda/CorreaCES20,DBLP:conf/ec/CorreaDFS19,DBLP:conf/soda/KaplanNR20,DBLP:conf/innovations/RubinsteinWW20}). In a model popularized by Kaplan et al. \cite{DBLP:conf/soda/KaplanNR20}, for example, an adversary chooses $k+n$ numbers. A random subset of $k$ of these numbers are revealed to the decision maker (as samples) at the outset. Afterwards, the decision maker gets to inspect the remaining $n$ numbers in an online fashion, and in random order.

While equally well motivated fairly little is known about these sampling models for the secretary objective. The only exemption is an elegant paper by Correa et al.~\cite{CorreaEtAl21}, which studies a variant of the Kaplan et al.~model with an additional independence assumption that serves to increase the mathematical tractability of the problem. 
In this variant of the model, an adversary chooses $n$ numbers, and each number is marked independently with probability $p$ as a sample. In this setting, they show that the optimal policy for any $n$ and $p$ is a threshold policy, obtain closed form solutions for thresholds when $n \rightarrow \infty$, and use this to recover the optimal approximation guarantees of the classic no advice model of $1/e$ (when $p \rightarrow 0$) and of the known distributions setting of $\approx 0.58$ (when $p \rightarrow 1$).

A different kind of information augmentation occurs frequently in real job markets in the form of recommendation letters. Suppose each candidate comes with a letter that makes claims about the candidate being the best in the pool. Obviously if the recommender is known to be 100\% accurate, the problem becomes easy. However, if the recommender can be wrong with some probability, the choice of the optimal policy is less obvious. 

All of these scenarios---assumptions on random arrival,  availability of quality scores, and recommendation letters---can be seen as a kind of ``advice'' given to the algorithm. In this work we take a general view of this problem, and explore secretary problems with advice.

\subsection{Our Contribution}

We unify disparate information augmentation settings for the secretary problem into a single framework that is rich enough to capture the classic random arrival model~\cite{dynkin1963optimum}, the Gilbert-Mosteller i.i.d. setting~\cite{gilbert2006recognizing}, its Markovian generalizations~\cite{allaart2010,DuToitEtAl2009,HLYNKA1988,YamEtAl2009}, and the recently introduced sample-based variants~\cite{CorreaEtAl21,DBLP:conf/soda/KaplanNR20}.  

In all of these settings, previous work painstakingly showed existence and optimality of threshold policies, which essentially specify the minimum hiring bar for every time step. We identify structural properties of the advice that explain the optimality of such policies as well as give an algorithm for quickly computing the optimum thresholds. 

To demonstrate the utility of the framework we investigate a new machine-learning inspired advice for this problem, prove that the optimal policy is a threshold policy, and give tight bounds on its performance. 

\paragraph{Setup}
Consider a setting where each secretary $i$ has a rank $r_i$. In the generic case without ties, the ranks form a permutation of $[n] = \{1,\dots,n\}$; we assume that higher is better so that $r_i = n$ is the secretary we seek to hire. 
We pair each secretary with an abstract signal $s_i$ from some signal space $\mathcal{S}$, and assume that ranks and signals are drawn from a known joint distribution $F$.

The crux of the problem is that although the algorithm knows this distribution, it cannot observe the rank of the arriving secretary directly. Instead, it can observe  the relative order of the secretaries that have arrived so far, along with their signals, and use this information to deduce the likelihood of the current secretary being the best.

\paragraph{Techniques} 
We identify two properties of signalling schemes, {\em Non-Filtering (NF)} and {\em History-Irrelevance (HI)}, that allow us to write down a factor revealing linear program that quantifies the performance of the best policy. 
The two properties restrict the joint distribution, $F$,  on the signals and the ranks. Roughly speaking, Non-Filtering implies that future signals do not change the information state available to the algorithm when processing a particular candidate. History-Irrelevance states that, conditioned on the current candidate being best so far, the probability of the candidate being globally optimum is independent of previously seen signals. We describe these precisely in Section \ref{sec:signal_properties}. 

We show the factor revealing LP and its dual in Figure~\ref{fig:lps}. The linear program is a generalization of the one proposed by Buchbinder et al.~\cite{buchbinder2010secretary} for the classic secretary program, with additional considerations to capture the effect of the signals. 
\begin{figure}
$$\begin{aligned}
& \max \sum_i \sum_s z(i,s) \cdot a(i,s) \text{ s.t. } \\
& z(i,s) \leq 1 - \sum_{j<i} \sum_{s'} z(j,s') \cdot c(i,s,j,s')\\
& z(i,s) \geq 0
\end{aligned} \qquad \left| \qquad \begin{aligned}
& \min \sum_i \sum_s u(i,s) \text{ s.t. } \\
& u(i,s) + \sum_{j>i} \sum_{s'}  u(j,s') \cdot c(j,s',i,s) \geq a(i,s) \\
& u(i,s) \geq 0
\end{aligned} \right.$$
\caption{Factor revealing LP for secretaries with advice and its dual.}\label{fig:lps}
\end{figure}
In the primal, the variable $z(i,s)$ denotes the probability of accepting a candidate with signal $s$ at time $i$ provided that it is the best candidate seen thus far. 
The coefficients $a(i,s)$ and $c(i,s,j,s')$ depend only on the joint distribution, $F$. In particular, $a(i,s)$ is the probability of seeing the top rank element in position $i$ with signal $s_i$, and $c(i,s,j,s')$ is the probability that the candidate in position $j$ with signal $s'$ is the best candidate thus far, given that candidate in position $i > j$ with signal $s$ is also locally optimum. 

Next, we characterize the policies captured by our linear program formulation. Here we show that under the NF and HI restrictions any stopping policy is captured by the primal, and any solution to the primal can be converted to a stopping rule.

\begin{propx}[Restatement of Proposition \ref{prop:policy_to_variable}]
For any policy for the secretary problem with a signaling scheme satisfying NF and HI there is a set of values $z(i,s) \geq 0$  such that the objective of the factor revealing linear program corresponds to the probability that the highest ranked secretary is selected.
\end{propx}

The converse is also true:

\begin{propx}[Restatement of Proposition \ref{prop:variable_to_policy}]
For any signaling scheme satisfying NF and HI, given feasible values $z(i,s) \geq 0$ there is a policy for the secretary problem which picks the highest ranked secretary with probability equal to the objective of the factor revealing linear program.
\end{propx}

We delve deeper to understand the kinds of optimal policies generated by the LP. Let {\em memoryless} policies be those that make the hiring decision at time $i$ based only on the signal of the secretary $i$, and {\em not} on the signals of previous candidates. Memoryless policies are a natural class, since their implementation does not require tracking previously seen signals, and can be computed in constant space. 

We prove that any policy captured by the LP in Figure \ref{fig:lps} is a {\em memoryless} policy. 

\begin{thmx}[Restatement of Theorem \ref{thm:simple_policy}]
If a signaling structure satisfies Non-Filtering and History-Irrelevance, then the optimal signaling policy is a memoryless policy.
\end{thmx}

An important subclass of memoryless policies are {\em threshold} policies. These policies associate an earliest acceptance time for every signal, and accept the first locally optimum candidate whose signal passes the test. 
 We characterize algorithmically when threshold policies are optimal under HI and NF: They are optimal precisely when the dual to the linear program can be solved optimally by the greedy (backward induction) algorithm and satisfies a natural monotonicity assumption.
 
 \begin{propx}[Restatement of Proposition \ref{prop:greedy_thresholds}]\label{propx:greedy_thresholds}
Assume NF and HI. If the greedy solution $u(i,s)$ for $i \in N$ and $s \in \mathcal{S}$ constructed via backwards induction solution is an optimal solution to the dual LP and for all $s \in S$, $u(i,s)$ is non-decreasing in $i$, then the optimal policy is a threshold policy.
\end{propx}

and

\begin{propx}[Restatement of Proposition \ref{prop:thresholds_greedy}]
Assume NF and HI. If the optimal policy is a threshold policy, then the greedy backwards induction solution is optimal for the dual LP.
\end{propx}

We also show that the monotonicity assumption in Proposition~\ref{propx:greedy_thresholds} is required: there are cases where the greedy algorithm leads to an optimal solution, but the solution violates monotonicty, and the optimal policy is not a threshold policy.

A key advantage of our approach is that it allows us to characterize the family of optimal policies fairly easily, in contrast with previous work where ad hoc lower bounds techniques were required. 
Furthermore, our factor-revealing linear program allows us to derive explicitly optimal policies and their success probability for several problems. We can obtain such results by a careful backward induction argument on the dual of our factor-revealing linear program. For example, in this way we can recover the optimal policy for the Gilbert and Mosteller setting, and for our binary setting with advice. 

\paragraph{Applications}

We highlight some of the results that can be derived using our framework. 

First we focus on secretary algorithms for the sampling model of Kaplan et al.~\cite{DBLP:conf/soda/KaplanNR20}. In this model an adversary chooses $n+k$ numbers, a random subset of $k$ of these numbers are presented to the algorithm as {\em samples} at the outset, the remaining $n$ numbers are presented to the algorithm one-by-one, in random order. This model smoothly interpolates between the classic secretary problem with no advice ($k = 0$) and the known distributions setting of Gilbert and Mosteller ($k \rightarrow \infty$).

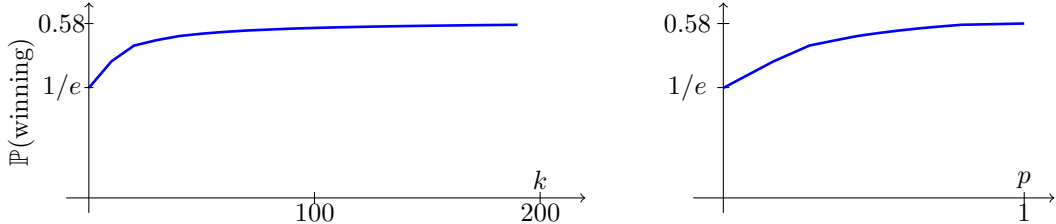
\begin{figure}[t]
\centering
\begin{tikzpicture}[xscale=3, yscale=4]
\draw[line width=1,  color=blue] (0.000000,0.365788)--(0.100000,0.454104)--(0.200000,0.506639)--(0.300000,0.524683)--(0.400000,0.538279)--(0.500000,0.546238)--(0.600000,0.552113)--(0.700000,0.556753)--(0.800000,0.560020)--(0.900000,0.563082)--(1.000000,0.565355)--(1.100000,0.567278)--(1.200000,0.568901)--(1.300000,0.570216)--(1.400000,0.571536)--(1.500000,0.572558)--(1.600000,0.573500)--(1.700000,0.574373)--(1.800000,0.575100)--(1.900000,0.575793);
\draw[->] (-.1,0) -- (2.2,0);
\draw[->] (0,-0.05) -- (0,0.65);
\draw (-.02,.367) -- (.02,.367);
\node at (-.12,.367) {$1/e$};
\draw (-.02,.58) -- (.02,.58);
\node at (-.12,.58) {$0.58$};

\node at (2,-.05) {$200$};
\draw (2,-.02) -- (2,.02);
\node at (1,-.05) {$100$};
\draw (1,-.02) -- (1,.02);
\node at (2,.06) {$k$};

\node[rotate=90] at (-.3,.32) {$\P(\text{winning})$};

\begin{scope}[xscale=4/3, xshift=60]
\draw[line width=1,  color=blue] (0.000000,0.365788)--(0.166667,0.454104)--(0.285714,0.506639)--(0.375000,0.524683)--(0.444444,0.538279)--(0.500000,0.546238)--(0.545455,0.552113)--(0.583333,0.556753)--(0.615385,0.560020)--(0.642857,0.563082)--(0.666667,0.565355)--(0.687500,0.567278)--(0.705882,0.568901)--(0.722222,0.570216)--(0.736842,0.571536)--(0.750000,0.572558)--(0.761905,0.573500)--(0.772727,0.574373)--(0.782609,0.575100)--(0.791667,0.575793)--(1,0.58);
\draw[->] (-.1,0) -- (1.1,0);
\draw[->] (0,-0.05) -- (0,0.65);
\draw (-.02,.367) -- (.02,.367);
\node at (-.12,.367) {$1/e$};
\draw (-.02,.58) -- (.02,.58);
\node at (-.12,.58) {$0.58$};
\node at (1,-.05) {$1$};
\draw (1,-.02) -- (1,.02);

\node at (1,.06) {$p$};
\end{scope}
\end{tikzpicture}
\caption{Performance of the optimal policy for secretary with samples. Both plots are for $n = 20$. On the left, the $x$-axis shows $k$ and the $y$-axis shows the probability of winning. On the right, the $x$-axis is $p = k/(k+n)$ and the $y$-axis is probability of winning.}
\label{fig:samples}
\end{figure}

In Section \ref{sec:samples} we show how to capture this problem within our LP approach; as a consequence we obtain an LP formulation for the Gilbert-Mosteller model. We use the dual to show the optimality of thresholding policies for all number of samples $k$ and all numbers of candidates $n$. The dual then has the interpretation that the $u(i,s)$ variables are the contribution to the optimal winning probability from secretary $i$ when seeing signal $s$. The dual also gives rise to an efficient poly-time algorithm for computing the optimal policy and the corresponding winning probability. See Figure~\ref{fig:samples} for an exemplary set of these bounds.

Furthermore, we obtain exact solutions to the dual for all $n$ and $k$, and thus the thresholds used by the optimal policy. These can be used to derive exact analytic expressions of the optimal winning probability in the asymptotic regime when $n \rightarrow \infty$. We do this for the Gilbert-Mosteller setting, and thus recover the aforementioned analytic expression of the winning probability.

Our results for this setting strengthen and extend the results of both \cite{CorreaEtAl21} and \cite{gilbert2006recognizing}. We use our framework to give an LP-duality based proof of the optimality of thresholding for all values of $n$ and $k$ (without the simplifying independence assumption of \cite{CorreaEtAl21}), we obtain an efficient (poly-time) algorithm for computing the optimal policy and winning probability for all values of $n$ and $k$, and derive closed formulas for the thresholds used by the optimal policy in both the asymptotic and non-asymptotic regime.

We then give an example of a new kind of advice that can be easily analyzed in this framework. Suppose each secretary arrives with a binary signal indicating whether the candidate is globally optimum, but the signal is incorrect with some probability $1-p$. Anecdotally such a signal can model recommendation letter writers that claim that a candidate is ``best in their class.''

Formally, we consider the random order arrival model, and restrict the signal space to $\mathcal{S} = \{Y, N\}$.  For the best candidate the signal is $Y$ with probability $p$; for each other candidate the signal is $N$ also with probability $p$. In Section \ref{sec:binary} we consider the extended case where the probability of error is different for two classes, capturing potential false positive and false negative trade-offs in a real life classifier.  

We show that the optimum solution in this setting is a threshold policy with two thresholds: 
$$t_Y \approx n \left(\frac{1}{p}-1\right)^{\nicefrac{1}{p}} e^{1 - \nicefrac{1}{p}} \qquad \text{and} \qquad t_N \approx n e^{1 - \nicefrac{1}{p}}.$$ 
In other words, the optimum policy waits the first $t_Y$ steps; then accepts any candidate with a $Y$ signal that is better than all previous candidates until time $t_N$, and then accepts any candidate that is better than all already rejected candidates.  Note that when $p = \nicefrac{1}{2}$, and thus the signal provides no additional information, we recover Dynkin's classic policy. However, as p grows, the thresholds $t_Y$ and $t_N$ diverge, and the competitive ratio grows to $\approx (1/p - 1)^
{(1/p - 1)} \cdot e^{(1 - 1/p)}$. 

\begin{figure}[t]
\centering
\begin{tikzpicture}[xscale=10, yscale=4]
\draw[line width=1,  color=blue] (0.5, 0.37104277871264424)--
 (0.55555555555555558, 0.37910695541424322)--
 (0.61111111111111116, 0.4003136327102359)--
 (0.66666666666666663, 0.43244764221685245)--
 (0.72222222222222221, 0.47513807686812032)--
 (0.77777777777777779, 0.52935023651537838)--
 (0.83333333333333326, 0.59758322965465616)--
 (0.88888888888888884, 0.68502240756169819)--
 (0.94444444444444442, 0.80291350237634862)--
 (1.0, 0.99999999999998879);
\draw[->] (0.45,0) -- (1.05,0);
\draw[->] (0.5,-0.05) -- (0.5,1.05);
\node at (.75,-.05) {$p$};
\node at (.53,-.05) {$0.5$};
\node at (1,-.05) {$1$};
\node at (.46,.367) {$1/e$};
\node at (.48,.03) {$0$};
\node at (.48,1) {$1$};
\node[rotate=90] at (.4,.5) {$\text{Competitive Ratio}$};

\begin{scope}[xshift=23]
\draw[line width=1,  color=blue] (0.5, 0.36787944117144233) -- (0.55555555555555558, 0.4493289641172217) -- (0.61111111111111116, 0.5292133415000504) -- (0.66666666666666663, 0.6065306597126334) -- (0.72222222222222221, 0.6807123983233854) -- (0.77777777777777779, 0.751477293075286) -- (0.83333333333333326, 0.8187307530779817) -- (0.88888888888888884, 0.8824969025845955) -- (0.94444444444444442, 0.9428731438548749) -- (1.0, 1.0);
\draw[line width=1,  color=red] 
(0.5, 0.36787944117144233) -- (0.55555555555555558, 0.30069512768373236) -- (0.61111111111111116, 0.25259303137532102) -- (0.66666666666666663, 0.21444097124017678) -- (0.72222222222222221, 0.18129472962122328) -- (0.77777777777777779, 0.15010660595931891) -- (0.83333333333333326, 0.11867987997168922) -- (0.88888888888888884, 0.085062267284683801) -- (0.94444444444444442, 0.046948854029558049) -- (1.0, 0.0);

\draw[->] (0.45,0) -- (1.05,0);
\draw[->] (0.5,-0.05) -- (0.5,1.05);
\node at (.75,-.05) {$p$};
\node at (.53,-.05) {$0.5$};
\node at (1,-.05) {$1$};
\node at (.46,.367) {$n/e$};
\node at (.48,.03) {$0$};
\node at (.48,1) {$n$};
\node at (1,.94) {$t_N$};
\node at (1,0.07) {$t_Y$};
\end{scope}
\end{tikzpicture}
\caption{Left plot shows the performance of the optimal secretary with advice policy as a function of $p$. The right plot shows the thresholds $t_Y$ and $t_N$ as a function of $p$.}
\label{fig:aproximation_plot}
\end{figure}
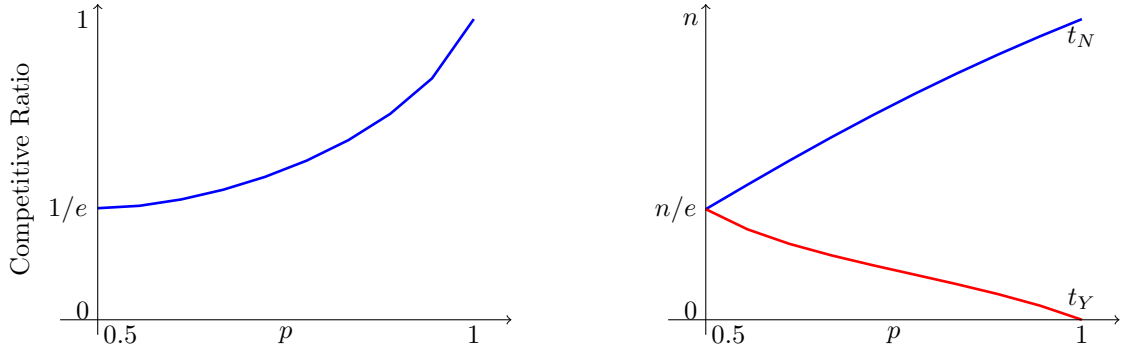

We plot both the growth of the competitive ratio as well as the two threshold $t_Y$ and $t_N$ in Figure \ref{fig:aproximation_plot}. 

\subsection{Related Work}

Our work is closely related to three broad active research directions. First, our problem formulation  is inspired and aligned with the nascent area of algorithms with machine-learned advice. Second, our technique can be seen as a unifying framework that emphasizes the role of advice in the secretary and prophets literature. Finally, on a technical level, our work is related to the literature using factor revealing LPs in approximation and online algorithms. In the following we briefly review the most related papers in those three areas.

\paragraph{Algorithms with ML Advice} 
Traditionally, the design and analysis of algorithms has focused on provable guarantees for worst-case inputs. A growing body of work explores how ``machine learned advice'' can be leveraged.
Thanks to its practical applications %and its theoretical depth 
several problems have been studied through this lens. 
Examples range from building better data structures~\cite{kraska2018case, mitzenmacher2018model}, to improved competitive and approximation ratios for several online tasks~\cite{lattanzi2020online, lykouris2018competitive,munoz2017revenue, mitzenmacher2020queues, purohit2018improving, rohatgi2020near}, to cases where advice has been used to speed-up algorithms~\cite{alabi2019learning, balcan2018learning} or to reduce their space complexity~\cite{hsu2018learning}. Our work can be seen as a formalization of the classic secretary problem in this general framework.

\paragraph{Secretaries and Prophets Literature} The secretary problem is one of the most studied problems in online optimization. The classic formulation of the problem, introduced by Dynkin~\cite{dynkin1963optimum}, 
succeeds with probability approaching $1/e$. The guarantee of $1/e$ is known to be best possible, even when the values of the secretaries are drawn from an unknown distribution (e.g., \cite{Ferguson89}).

The same objective has also been studied in the setting where the $n$ candidates are i.i.d.~draws from a known distribution (which can w.l.o.g.~be the uniform distribution on [0,1]) \cite{BerezovskiyG84,gilbert2006recognizing,gnedin96,Samuelson82}. The optimal algorithm sets a decreasing sequence of thresholds, and it can be shown that it accepts the best secretary with probability at least $\approx 0.58$.
A recent paper by Esfandiari et al.~\cite{EsfandiariHLM20} has considered a non-i.i.d variant of this problem, and shows how to obtain a $1/e$ approximation with a single threshold; and provides an example of non-i.i.d. distributions where this is best possible.

Another popular generalization of the Gilbert-Mosteller model (because of its application in finance) are random walk models. An early example is \cite{HLYNKA1988}. More recent results include \cite{allaart2010,DuToitEtAl2009,YamEtAl2009}. The main take away from this line of work is known as the ``bang-bang principle'': if the random walk is balanced then you might as well stop immediately, if it's biased upward you should wait until the end, if it's biased downwards you should accept immediately.

Another interesting line of work in this area studies the secretary and prophet problems in the presence of a limited number of samples~\cite{CorreaEtAl21,DBLP:conf/soda/CorreaCES20,DBLP:conf/ec/CorreaDFS19,  DBLP:conf/soda/KaplanNR20, DBLP:conf/innovations/RubinsteinWW20}. Most relevant in this context are the Kaplan et al.~paper \cite{DBLP:conf/soda/KaplanNR20}, which is the model we adopt here, and the paper by Correa et al.~\cite{CorreaEtAl21} as it is the only prior work that looks at the secretary objective.

Our work offers a unifying lens that captures all these problems as secretary problems with advice; and extends the known LP formulation for the classic secretary problem to all of the other problems. It in particular enables structural insights about the form the optimal policy takes, e.g, when and why backward induction yields optimal solutions.

\paragraph{Factor Revealing LPs}
Factor revealing LPs have been used in a number of algorithmic analyses. They were introduced in the context of designing approximation algorithms for the facility location problem, in conjunction with the dual fitting technique \cite{JainEtAl03}.
The technique has been extended to strongly factor revealing LPs by \cite{MahdianY11}, who used it to analyze the KVV ranking algorithm for bipartite matching in the random order model. Another variant called tradeoff revealing LPs was introduced in \cite{MehtaEtAl15} to analyze a greedy algorithm for the the AdWords problem.

In the context of secretary problems, there are two main precursors: The first is
%% there is also a MOR'14 version
\cite{buchbinder2010secretary}, which describes a LP that recovers the optimal $1/e$ approximation guarantee for the classic secretary problem. The second one is \cite{ChanEtAl15}, which extends this formulation to the $(j,k)$-secretary problem. In this variant of the problem the algorithm is allowed to retain $j$ elements and the goal is to maximize the expected number of elements that are among the $k$ best secretaries.

Our LP formulation is inspired by \cite{buchbinder2010secretary}, but is much more general, and in particular enables---for the first time---a unified treatment of the two classic secretary problems through the lens of LPs.

\medskip

\paragraph{Additional Related Work} 
In parallel to this work, Antonianidis et al.~\cite{AntoniadisEtAl20} have considered the value maximization variant of the secretary problem with advice (so a different objective than we consider here). The techniques and results of that paper are very different from those in this paper.

They consider, for example, the single choice problem, and as advice the anticipated quality $\rho^*$ of the best secretary. They seek bounds that are at least $\alpha > 1/e$ when the advice is accurate (as measured by the Euclidean distance) and at least $\beta < 1/e$ when the advice is inaccurate.
They obtain qualitatively similar results for more general combinatorial allocation problems, such as bipartite matching.

They do not characterize optimal policies and how their performance decays as the advice gets worse, and they also don't provide a general framework for studying different forms of advice.

%% SECTION 2
\section{Model of Secretaries with Advice}\label{sec:moodel}

In the original secretary problem, the algorithm's goal is hiring the best secretary from a set of $n$ candidates. There is a total order on the candidates which is not known in advance.
Candidates arrive in random order, and upon arrival, the algorithm is able compare the candidate with each of the previously seen options. The algorithm must then irrevocably decide to either hire that secretary or pass, in which case that secretary is no longer available.

Mathematically, we can describe the problem as follows: Let $(r_1, \dots, r_n)$ be a permutation of $[n] = \{1, \dots, n\}$.
We refer to $r_i$ as the \emph{rank} of the $i$-th arriving secretary. We say that $i \succeq j$ (i.e. secretary $i$ is at least as good as $j$) whenever $r_i \geq r_j$. Note that the best secretary has value $n$. The algorithm has no access to the ranks. Instead, at time $i$,  it can only see the relative comparisons $i \succeq j$ and/or $i \preceq j$ for $j < i$.
The goal is to maximize the probability with which we stop at the maximum. 

While we described the model without the possibility of ties, in some of our applications it will be natural to allow for ties. In that case we will assume that $(r_1, \hdots, r_n)$ is a vector with $r_i \in [n]$, and we will require that $r_i = n$ for at least one secretary $i$. Our goal will then be to stop at an $i$ such that $r_i = n$ (of which there may be more than one).

We will use $\preceq_{1..i}$ to represent the partial order induced on the first $i$ elements. 

\paragraph{Best-so-far Event}
We define a probabilistic event that will play a key role in the analysis and definitions below. Let the \emph{best-so-far} event $T_i$ be:
\begin{equation}\label{eq:ti}
T_i = \{i \succeq j; \forall j < i\}
\end{equation}
It is important to observe that since the goal of the algorithm is to pick the best secretary, one can assume without loss of optimality  that the algorithm only picks secretary $i$ if it is the best-so-far.

\paragraph{Advice} We augment the secretary problem with an extra signal $s_i$ for each arriving secretary. Let $\S$ be the space of signals. We assume that $(r_1, r_2, \hdots, r_n, s_1, \hdots, s_n)$ is drawn from a known joint distribution. In the no ties case, $r_1, \dots, r_n$ will just be a permutation of $[n]$ and $s_i \in \mathcal{S}$ for all $i$. With ties, we require that $r_i \in [n]$, $s_i \in \S$, and $r_i=n$ for some $i$.

In each period $i$ the algorithm observes both the signal $s_i$ and the relative comparisons $i \succeq j$ or $i \preceq j$ for $j < i$. %maximize $\P[r_\tau = n]$ where $\tau$ is the \emph{stopping time} induced by the algorithm. 
The algorithm knows the joint distribution of ranks and signals, but it cannot observe ranks directly, it is limited to computing induced ranks.

The algorithm decides in each step $i$ whether to stop or proceed. 
As before, the goal of the algorithm is to select the best candidate. 
Note that if there are two secretaries with the top rank, we can pick either one.

\subsection{Examples of Secretary Problems with Advice}

It is useful to keep some concrete examples in mind:

\begin{example}[Secretaries without advice~\cite{dynkin1963optimum}]\label{ex:no_advice}
If $\S = \{0\}$ and $(r_1, \hdots, r_n)$ are distributed as a random permutation, we are back at the original secretary problem. The best optimal strategy for this problem (Dynkin's algorithm) finds the optimal secretary with probability $1/e \approx 0.37$.
\end{example}

\begin{example}[Gilbert-Mosteller~\cite{gilbert2006recognizing}]\label{ex:gm}
Consider a fixed known distribution $F$ over the real numbers. Let $\S = \R$, $s_i$ is an independent sample from $F$ and $r_i$ represent the ranks induced by $s_i$, i.e., $r_i = k$ if $s_i$ is the $k$-th largest value of among $(s_1, \hdots, s_n)$. This stochastic version of the secretary problem is studied by Gilbert and Mosteller who show that with this extra information the algorithm can hire the best secretary with probability $\approx 0.58$.

A non-i.i.d.~version of Gilbert-Mosteller model was studied by Esfandiari et al.~\cite{EsfandiariHLM20} who show that when $s_i \sim F_i$ then it is possible to choose the optimal secretary with probability $1/e$; and this is best possible in the worst case. (Note that this does not follow from Dynkin's algorithm since the ranking induced by the random draws is no longer uniform random.)
\end{example}

\begin{example}[Markovian stopping]\label{ex:ms}
A generalization of the Gilbert-Mosteller setting is the following Markovian stopping problem: consider a Markov chain on space $\S$, i.e., a stochastic process $s_1, \hdots, s_n$ where $\P[s_i \mid s_1, \hdots, s_{i-1}] = \P[s_i \mid s_{i-1}]$~and let $r_i$ be the ranks induced by signals assuming there is a total ordering defined on $\S$.

Markovian stopping problems are popular in finance, where they serve as proxies for investment problems. Hlynka and Sheahan \cite{HLYNKA1988}, for example, study a simple ``symmetric'' random walk. The process starts on day zero with a reward of zero. Then on each of $n$ days, with equal probability, either the reward is increased by one or it is decreased by one. The goal is to maximize the probability with which the process is stopped at the maximum reward of all days. They show that all strategies that skip a fixed number of $t \geq 0$ days, and then accept the first reward from day $t+1$ onwards that is the highest so far actually achieve the exact same winning probability.

Subsequent work has identified this as the indifference case of what has become to be known as the ``bang bang principle'' \cite{allaart2010,DuToitEtAl2009,YamEtAl2009}: In a random walk that is started at zero and in which the probability to move up by one is $p$ and the probability to move down by one is $1-p$ it is best to stop immediately when $p < 1/2$ and to wait until the end when $p > 1/2$.
\end{example}

\begin{example}[Secretaries with Samples~\cite{CorreaEtAl21,DBLP:conf/soda/KaplanNR20}]\label{ex:samples} 
An adversary writes down $n+k$ numbers. A random subset of size $k$ is chosen and revealed to the algorithm as samples. Afterwards, the remaining $n$ numbers are presented to the algorithm in an online fashion, in random order.

The algorithm can observe the relative order of all secretaries it has seen so far. So in addition to observing the best so far event $T_i$, the algorithm learns about the relative rank of the current secretary among the samples. The signal space is thus $\mathcal{S} = \{0, \dots, k\}$, where $s_i = j$ means that $j$ of the samples are worse than the current secretary.

The rank $r_i$ of a hirable secretary is its relative rank among the $n$ hirable secretaries. So $r_i = n$ is the rank of the best secretary, and $r_i = 1$ is the rank of the worst secretary.
\end{example}

\begin{example}[Binary Classifier]\label{ex:yes_no}
A natural ML-advice model is a classifier that given a secretary predicts whether it is the best secretary or not. The input to the machine learned system is a candidate with all of their features,  and the output is a binary classification: Y(es) or N(o). 

In ML it's common practice to evaluate the quality of a binary classifier using the following four metrics \emph{accuracy, precision, recall, and specificity}. It will be more convenient for us to express our results in terms of \emph{recall} (a.k.a.~\emph{sensitivity}) and \emph{specificity}. Such metrics are depicted in Figure \ref{fig:recall_specificity} as a function of True Positive (TP), True Negative (TN), False Positive (FP) and False Negative (FN).

\begin{figure}[h]
\centering
\begin{tikzpicture}[scale=.6]
\draw (0,0) rectangle (4,4);
\draw (0,2) -- (4,2);
\draw (2,0) -- (2,4);
\node at (1,1) {FN};
\node at (1,3) {TP};
\node at (3,1) {TN};
\node at (3,3) {FP};
\node at (2,5.1) {Actual};
\node at (1,4.4) {Y};
\node at (3,4.4) {N};
\node[rotate=90] at (-1.1,2) {Predicted};
\node at (-.4,3) {Y};
\node at (-.4,1) {N};
\node[anchor=west] at (6,3) {Recall: $p = \frac{TP}{TP + FN}$};
\node[anchor=west] at (6,1) {Specificity: $p' = \frac{TN}{FP + TN}$};
\end{tikzpicture}
\caption{Parameters of the binary classifier}
\label{fig:recall_specificity}
\end{figure}

This translates to a signal space $\S = \{Y, N\}$ and two parameters $p,p' \in [\frac{1}{2},1]$ corresponding to recall and specificity respectively. We will assume that $(r_1, \hdots, r_n)$ is a random permutation and that:
$$\begin{aligned}
& \P[s_i = Y \mid r_i = n] = p & \qquad & \P[s_i = N \mid r_i = n] = 1-p\\
& \P[s_i = Y \mid r_i = r] = 1-p' & \qquad & \P[s_i = N \mid r_i = r] = p'
\end{aligned}$$
for every $r < n$. Essentially the advice suggest to hire the best secretary with probability $p$ and suggest to hire any other secretary with probability $1-p'$. When $p=p'=\frac{1}{2}$ we are back in the original secretary problem since the signal is clearly useless. When $p=p'=1$ the ML model is perfect and we can hire the optimal secretary by simply following its suggestion, i.e., hiring whenever it says $Y$ and not hiring otherwise.
\end{example}

\subsection{Signal Structure Properties}
\label{sec:signal_properties}

The joint distribution in all of the examples we described above has two properties that will be key to our analysis: {\em Non-Filtering} and {\em History Irrelevance}. As we note below in {Example \ref{ex:no_nfhi}} these are not universal and rule out certain types of advice. 

\begin{definition}[Non-Filtering]\label{def:nf}
We say that a signaling structure satisfies Non-Filtering (NF) if given any $j<i$ we have:
$$\P[\preceq_{1..j}, s_1, \hdots, s_{j} \mid T_j, s_j] = \P[\preceq_{1..j}, s_1, \hdots, s_{j} \mid T_j, s_j, T_i, s_i]$$
where $T_i$ is the best-so-far event defined in equation \eqref{eq:ti}.
\end{definition}

The property essentially means that if at both $i$ and $j$ we see secretaries that are the best-so-far the information we see later in $i$ does not affect the conditional distribution of information the algorithm gets at time $j$. 

Note that it holds in Examples \ref{ex:gm}, \ref{ex:ms} simply by the Markovian property: conditioned on $T_j, s_j$, we have that $(\preceq_{1..j}, s_1, \hdots, s_{j})$ and $(s_j, s_{j+1}, \hdots, s_n)$ are independent. Since $s_i, T_i$ is a function of $(s_j, s_{j+1}, \hdots, s_n)$, property NF automatically follows. A similar, but more delicate argument which we defer to Section \ref{sec:samples}, can be used to argue that Example \ref{ex:samples} also satisfies NF. 
In Example \ref{ex:yes_no} if $\P'(N) = p'$ and $\P'(Y) = 1-p'$ we have:
$$\P[\preceq_{1..j}, s_1, \hdots, s_{j} \mid T_j, s_j, T_i, s_i ] =  \frac{1}{(j-1)!} \prod_{t=1}^{j-1} \P'( s_t)$$
since after we know that $j$ is the best so far, each element before $j$ cannot be the top ranked element and hence must have its signals drawn from $\P'$.

Non-Filtering captures the effect of future arrivals on the information state available to the algorithm. In contrast, history irrelevance imposes structure on the relationship between signals previously observed. 

\begin{definition}[History-Irrelevance]\label{def:hi}
We say that a signaling structure satisfies History-Irrelevance (HI) if conditioned on $(T_i, s_i)$ the variable $(\prec_{1..i}, s_1, \hdots, s_i)$ representing the information available at round $i$ and the event $r_i = n$ are independent.
\end{definition}

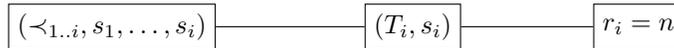
\begin{figure}[h]
\centering
\begin{tikzpicture}[every node/.style={draw, rectangle}, minimum size = .6cm]
\node (a) at (0,0) {$(\prec_{1..i}, s_1, \hdots, s_i)$};
\node (b) at (4,0) {$(T_i,s_i)$};
\node (c) at (7,0) {$r_i = n$};
\draw (a)--(b);
\draw (b)--(c);
\end{tikzpicture}
\caption{Graphical model representing property HI}
\label{fig:hi}
\end{figure}

In graphical model language, we can say that the variables and event above satisfy the graphical model in Figure~\ref{fig:hi}. It is useful to observe some implications of this fact. The first implication is that in order to determine the probability that each $i$ is the highest ranked secretary, it is enough to look at its signal and to whether it is the highest ranked so far:
\begin{equation}\label{eq:hi1}
\P[r_i = n \mid \prec_{1..i}, s_1, \hdots, s_i] = \P[r_i = n \mid T_i,  s_i]\end{equation}
A second implication is that if $X$ is an event that depends only on the the information available at period $i$ (i.e., depends only on $(\prec_{1..i}, s_1, \hdots, s_i)$) then:
\begin{equation}\label{eq:hi2}
\P[X \mid s_i, r_i = n] = \P[X \mid T_i,  s_i]\end{equation}
since $r_i=n$ implies $T_i$. 

Again it is simple to see that all the example discussed satisfy it. For Examples \ref{ex:gm} and \ref{ex:ms} it again follows from the same argument: $(\preceq_{1..j}, s_1, \hdots, s_{i})$ and $(s_i, s_{i+1}, \hdots, s_n)$ are conditionally independent given $T_i, s_i$ and $r_i = n$ depends only on $(s_i, s_{i+1}, \hdots, s_n)$ whenever $T_i$ holds. For Example \ref{ex:yes_no} we can explicitly compute the probability $\P[r_i=n\mid \preceq_{1..i}, s_1, \hdots, s_i]$ as follows: if $i$ is not the best so far, then this probability is zero. Subject to $T_i$ the probability is $\frac{p }{p  + (n-i)(1-p') }$ if $s_i = Y$ and $\frac{1-p}{1-p  + (n-i)p' }$ if $s_i = N$. Hence, conditioned on $(T_i, s_i)$ the probability of $r_i=n$ is independent of $(\preceq_{1..i}, s_1, \hdots, s_i)$. For Example \ref{ex:samples} we again defer the discussion to Section \ref{sec:samples}.

%% SECTION 3
\section{Memoryless Policies and LP Formulation}\label{sec:general_lp}

Non-Filtering (NF) and History-Irrelevance (HI) are \emph{natural} properties of a signalling scheme. In this section we investigate which policies are optimal under Non-Filtering (NF) and History-Irrelevance (HI). %properties.

\paragraph{Memoryless Policies} 
A generic policy is a map from the available information at period $i$, which consists of $(\preceq_{1..i}, s_1, \hdots, s_i)$ to a stopping probability. In general the optimal generic policy can be quite complicated, specially when there are lots of correlations among signals. An interesting class of policies are the one where the decision is based on the advice given for the current secretary only and where the decision does not depend on the relative order among the first $i-1$ elements, but instead depends only on whether $i$ is the best-so-far or not, i.e., whether we are in the event $T_i$. Putting it all together, we say that a \emph{memoryless policy} is a map that outputs the probability of stopping in each round $i$ conditioned on $T_i$ and $s_i$.

Memoryless policies are well-motivated and a natural assumption in ML applications. The resulting policies are simpler, and require less space. They offer increased privacy and may even be a legal necessity (e.g., when data protection laws regulate which data may be stored and for how long).

Under the following natural conditions (satisfied by Examples \ref{ex:no_advice}, \ref{ex:gm}, \ref{ex:ms}, \ref{ex:samples}, and \ref{ex:yes_no}) the optimal signaling policy is memoryless:

\begin{theorem}\label{thm:simple_policy} If a signaling structure satisfies Non-Filtering and History-Irrelevance, then the optimal signaling policy is a memoryless policy.
\end{theorem}

Even under these nice conditions, it is not trivial to show the theorem above. As one tries to modify a generic policy to make it be oblivious to the history, one needs also to account for probability of not having stopped at any given point which is complicated to track. 

\paragraph{LP Formulation and Proof of Theorem~\ref{thm:simple_policy}} The main tool we will use is a linear programming formulation for the secretary problem with advice subject to the NF and HI properties. Our formulation generalize the linear program of Buchbinder, Jain, and Singh \cite{buchbinder2010secretary} to our more challenging setting. 

Assume the signaling structure satisfies NF and HI and consider a generic policy mapping $(\preceq_{1..i}, s_1, \hdots, s_i)$ to a stopping probability for each $i$.  For each $i \in [n]$ and $s \in \S$, we define:
\begin{align*}
z(i,s) &= \P \left[ i \text{ is picked}\mid T_i, s_i = s \right]
\end{align*}
We furthermore define for each $i \in [n]$ and $s \in \S$, $a(i,s) = \P\left[r_i = n, s_i = s\right]$ and for each $i,j \in [n]$ with $i > j$ and $s,s' \in \S$, $c(i,s,j,s') = \P\left[T_j, s_j=s' \mid T_i, s_i = s\right]$.

We will show that if $z(i,s)$ is ``feasible'' (to be defined soon), then it contains all the relevant data to reconstruct the policy.

Which constraints must $z(i,s)$ satisfy in order for us to be able to recover a policy from it? Besides the trivial requirement that $z(i,s) \geq 0$ the other requirement we will ask is that there is enough probability left to choose $i$ when $i$ arrives. Even if $i$ is the top among the first $i$ elements, an earlier element may have been chosen preventing us from choosing $i$ later on. In other words, we have:
%$$\textcolor{violet}{\P\left[i \text{ is picked } \mid T_i \text{ and } s_i = s\right] +} 
$$\P\left[i  \text{ is picked}\mid T_i , s_i = s \right] + \sum_{j < i} \P\left[j  \text{ is picked}\mid T_i , s_i = s \right] \leq 1.$$
Since each $j < i$ is only picked in the event $T_j$ ($j$ is the best so far) then we can re-write the expression above using the law of total probability as:
$$\begin{aligned}& \P \left[ i \text{ is picked}\mid T_i, s_i = s \right] + \sum_{j < i} \sum_{s' \in \S} \P\left[j  \text{ is picked}\mid T_j, s_j = s',T_i, s_i = s \right] \P \left[ T_j, s_j = s' \mid T_i, s_i = s \right] \leq 1.\end{aligned}$$
Now we can apply property NF to argue that:
$$\P\left[j  \text{ is picked}\mid T_j, s_j = s', T_i, s_i = s\right] = \P\left[j  \text{ is picked}\mid T_j, s_j = s' \right].$$
Since the fact that $j$ is picked depends only on $(\preceq_{1..j}, s_1, \hdots, s_j)$ and the conditional distribution of $(\preceq_{1..j}, s_1, \hdots, s_j)$ is the same given $T_j,s_j$ or $T_j,s_j,T_i,s_i$ we obtain the last display equation. Substituting it above and replacing the definition of $z(i,s)$ and $c(i,s,j,s')$ we obtain:
\begin{equation}\label{eq:main_const}
%z(i,s) \leq 1- \sum_{j < i} \sum_{s' \in \S} z(j,s') \cdot \P \left[ T_j  \text{ and } s_j = s' \mid T_i \text{ and } s_i = s \right]\\
z(i,s) \leq 1- \sum_{j < i} \sum_{s' \in \S} z(j,s') \cdot c(i,s,j,s').
\end{equation}

It is important to note that the term $c(i,s,j,s') = \P \left[ T_j, s_j = s' \mid T_i, s_i = s \right]$ depends only on the joint distribution of $(r_1, \hdots, r_n, s_1, \hdots, s_n)$ and not on the policy itself.

We can also write the performance of the policy in terms of the $z(i,s)$ variables:
$$\begin{aligned}\Obj & = \sum_{i=1}^n  \E_{\preceq_{1..i}, s_1, \hdots, s_i} \left[ \P \left[ i \text{ is picked} \wedge r_i = n \mid \preceq_{1..i}, s_1, \hdots, s_i \right]\right] \\ & = \sum_{i=1}^n  \E_{\preceq_{1..i}, s_1, \hdots, s_i} \left[ \P \left[ i \text{ is picked} \mid \preceq_{1..i}, s_1, \hdots, s_i \right] \cdot \P \left[r_i = n \mid \preceq_{1..i}, s_1, \hdots, s_i  \right] \right] \\ & = \sum_{i=1}^n  \E_{\preceq_{1..i}, s_1, \hdots, s_i} \left[ \P \left[ i \text{ is picked} \mid \preceq_{1..i}, s_1, \hdots, s_i \right] \cdot \P \left[r_i = n \mid T_i,s_i  \right] \right]
\\ & = \sum_{i=1}^n  \E_{s_i} \left[ \P \left[ i \text{ is picked} \mid T_i, s_i \right] \cdot \P \left[r_i = n \mid s_i  \right] \right],\end{aligned} $$
where the first equality follows from independence, the second from HI (equation \eqref{eq:hi1}) and the third follows from the law of conditional probability.  Substituting $z(i,s)$ and $a(i,s)$ we get:
\begin{equation}\label{eq:obj}
    \Obj =  \sum_{i=1}^n \sum_{s \in \S} z(i,s) \cdot a(i,s).
\end{equation}
We note again that the term $a(i,s)=\P[r_i=n, s_i=s]$ depends only on the signaling structure and not on the policy itself. We showed the following statement:

\begin{proposition}\label{prop:policy_to_variable}
For any policy for the secretary problem with a signaling scheme satisfying NF and HI there is a set of values $z(i,s) \geq 0$ satisfying \eqref{eq:main_const} such that the objective in equation \eqref{eq:obj} corresponds to the probability that the highest ranked secretary is selected.
\end{proposition}

The converse is also true:

\begin{proposition}\label{prop:variable_to_policy}
Assume again that the signaling structure satisfies NF and HI.
Given values $z(i,s) \geq 0$ satisfying \eqref{eq:main_const} then there is a policy for the secretary problem which picks the highest ranked secretary with probability equal to the objective defined in \eqref{eq:obj}.
\end{proposition}

\begin{proof}
Consider the policy that upon seeing secretary $i$ with signal $s$ chooses that secretary with probability:
\begin{align}
q(i,s) = \frac{z(i,s)}{1- \sum_{j < i} \sum_{s' \in \S} z(j,s') \cdot \P \left[ T_j  \text{ and } s_j = s' \mid T_i \text{ and } s_i = s \right] } \label{eq:from-lp-to-policy}
\end{align}
if $i$ is the best-so-far (in other words, if $T_i$ happens) and zero otherwise. Now we need to argue that the probability that this policy chooses the highest ranked secretary is equal to the objective in equation \eqref{eq:obj}.

Before we do that, we show that under the reconstructed policy, the probability that we pick the $i$-th secretary conditioned on $T_i, s_i=s$ is indeed $z(i,s)$. We will show that recursively. Assume for now it is true for all $j<i$. Under the constructed policy the probability that we pick $i$ is the probability that we reach that step without picking any of the previous secretaries times the probability we choose $i$ at that step:
$$\begin{aligned} \P[i \text{ is picked} \mid T_i, s_i=s ] & = \P[\text{reach step } i \mid T_i, s_i =s] \cdot q(i,s)\\ & = \bigg( 1-\sum_{j<i} \P[j \text{ is picked} \mid T_i, s_i =s]\bigg) \cdot q(i,s).
\end{aligned}$$
We can now use the induction hypothesis to evaluate the probability that $j$ is picked:
$$\begin{aligned}\P[j \text{ is picked} \mid T_i, s_i = s] & = \sum_{s' \in \S} \P[j \text{ is picked} \mid T_j, s_j= s', T_i, s_i = s] \cdot \P[T_j, s_j= s' \mid T_i, s_i = s]\\ & = \sum_{s' \in \S} \P[j \text{ is picked} \mid T_j, s_j= s'] \cdot \P[T_j, s_j= s' \mid T_i, s_i = s] \\ & = \sum_{s' \in \S} z(j,s') \cdot \P[T_j, s_j= s' \mid T_i, s_i = s],
\end{aligned}$$
where the last equality follows from the induction hypothesis. Now taking the two previous display equations together and substituting the formula for $q(i,s)$ we obtain that:
$$\P[i \text{ is picked} \mid T_i, s_i=s ] = z(i,s).$$
Equipped with that we can now bound the performance of the policy:
$$\Alg = \sum_{i=1}^n \P[r_i = n] \cdot \left(\sum_{s \in \S} \P[s_i = s \mid r_i = n] \cdot \P[i \text{ is picked} \mid s_i = s, r_i = n] \right).$$
Since the probability that the algorithm picks $i$ depends only on $(\preceq_{1..i}, s_1, \hdots, s_i)$ we can use property HI (equation \eqref{eq:hi2}) to get that:
$$\P[i \text{ is picked} \mid s_i = s, r_i = n] = \P[i \text{ is picked} \mid T_i, s_i = s].$$
Putting it all together we get that:
$$\begin{aligned}
\Alg & = \sum_{i=1}^n \P[r_i = n] \cdot \left(\sum_{s \in \S} \P[s_i = s \mid r_i = n] \cdot z(i,s) \right)  =\Obj,
\end{aligned}$$
as claimed.
\end{proof}

We now can show the proof of Theorem \ref{thm:simple_policy} as a corollary:

\begin{proof}[Proof of Theorem \ref{thm:simple_policy}] Given any policy use Proposition \ref{prop:policy_to_variable} to obtain variables $z(i,s)$. Now use Proposition  \ref{prop:variable_to_policy} to convert this back to a policy. Observe that the policy we obtain from this transformation uses only  $s_i,T_i$ when deciding whether to stop. Hence, it is an memoryless policy.
\end{proof}

Since all of the examples in Section \ref{sec:moodel} satisfy the NF and HI conditions, Theorem \ref{thm:simple_policy} implies that they all have an LP formulation, and moreover, they all have optimal memoryless policies. 

We conclude this section with an example in which the signalling scheme does not satisfy NF and HI, and the optimal policy is not memoryless. 

\begin{example}
\label{ex:no_nfhi}
Fix some integer $m$ between $1$ and $n$ and consider two signals $\S = \{T,B\}$ such that $s_i = T$ if $r_i > m$ and $s_i = B$ otherwise. In other words, the signal indicates whether we are in the top (T) or bottom (B) of the distribution. This structure violates NF.

The optimal policy in that case is obvious: ignore all bottom elements and treat the top elements as a standard instance of the secretary problem with $n-m$ elements. Dynkin's policy on the reduced instance would say that we don't pick until we have seen at $(n-m)/e$ top elements and after that we pick the best so far. Such policy is not memoryless as it needs to remember how many top elements appeared up to a certain point.
\end{example}

%% SECTION 4
\section{Optimality of Threshold Policies}\label{sec:thresholds}

We showed that under NF and HI the optimal policy is memoryless. Here we investigate when the optimal strategy has the even simpler form of a threshold strategy. 
It will also be a good opportunity to study the structure of the dual LP. 

\paragraph{Threshold Policies} The optimal policy for the secretary problem without advice (Example~\ref{ex:no_advice}) is to wait until we have seen $n/e$ secretaries and then pick the first secretary that is the best so far. This is a special case of what we call a \emph{threshold policy}, which conceptually is just a further restriction of memoryless policies. Recall that we defined the \emph{best-so-far} event $T_i$ as:
$$
T_i = \{i \succeq j; \forall j < i\}.
$$
Now we say that a stopping time $\tau$ is a threshold policy if there is a threshold function $t : \S \rightarrow [n]$ such that:
$$\tau = \min\{i \text{ s.t. } T_i  \text{ and } i \geq t(s_i) \}.$$
Dynkin's algorithm for the secretary problem without advice has $t(\emptyset) = n/e$. It is useful to see what a threshold policy would look like for Yes/No advice (Example \ref{ex:yes_no}). A threshold policy in that case should specify threshold $t(N)$ and $t(Y)$. Assume for now that $t(Y) \leq t(N)$ (we will prove it should be the case later). Then the policy would say:
\begin{itemize}
    \item No item is picked for $i < t(Y)$.
    \item For items arriving $t(Y) \leq i < t(N)$ they are chosen if they are the best so far and the advice is Yes.
    \item For items arriving $t \geq t(N)$ they are chosen if they are the best so far regardless of the advice.
\end{itemize}
For $p=1$ the optimal policy is clearly a threshold policy with $t(Y) = 1$ and $t(N) = n$. For $p=\frac{1}{2}$ the optimal policy is the same as in Dynkin's algorithm: $t(Y)=t(N) = n/e$.

\paragraph{Characterization via Dual LP}
We will see that whether or not the optimal policy is a threshold policy is closely related to our ability to solve the \emph{dual of the LP} in the previous section with a greedy algorithm. 
More precisely, we will show that under NF and HI the optimal policy is a threshold policy if we can solve the dual with the greedy algorithm and the resulting solution satisfies a natural monotonicity assumption. This will be the case in both of our case studies. The reverse implication is always true: If the optimal policy is a threshold policy, then we can find it with the greedy algorithm.

To state our result more formally, let's first recall the LP from the previous section, and let's also derive its dual. Under NF and HI we have the following primal-dual pair:

$$\begin{aligned}
& \max \sum_i \sum_s z(i,s) \cdot a(i,s) \text{ s.t. } \\
& z(i,s) \leq 1 - \sum_{j<i} \sum_{s'} z(j,s') \cdot c(i,s,j,s')\\
& z(i,s) \geq 0
\end{aligned} \qquad \left| \qquad \begin{aligned}
& \min \sum_i \sum_s u(i,s) \text{ s.t. } \\
& u(i,s) + \sum_{j>i} \sum_{s'}  u(j,s') \cdot c(j,s',i,s) \geq a(i,s) \\
& u(i,s) \geq 0
\end{aligned} \right.$$\\
where $a(i,s)$ and $c(i,s,j,s')$ are coefficients in $[0,1]$ that depend only on the joint distribution of signals and ranks. In particular, $a(i,s) = \P[s_i = s , r_i = n]$ and $c(i,s,j,s') = \P[T_j, s_j= s' \mid T_i, s_i = s]$.

Furthermore note that since the coefficients $c$ are derived from a probability distribution then:
\begin{equation}\label{eq:c_prob}
    \sum_{s'} c(i,s,j,s') \leq 1.
\end{equation}

Consider solving the dual LP with the following greedy algorithm: Set $u(n,s) = a(n,s)$ for all $s$. Then for $i < n$ assuming that we have set $u(j,s')$ for $j > i$ and all $s'$, set
\begin{align}
    u(i,s) = \max\left\{0,a(i,s) - \sum_{j > i} \sum_{s'} u(j,s') \cdot c(j,s',i,s)\right\}. \label{eq:dual_dp}
\end{align}

It is clear that this leads to a feasible dual solution.

We show:

\begin{proposition}\label{prop:greedy_thresholds}
Assume NF and HI. If the greedy solution $u(i,s)$ for $i \in [n]$ and $s \in \mathcal{S}$ constructed via equation (\ref{eq:dual_dp}) is an optimal solution to the dual LP and for all $s \in S$, $u(i,s)$ is non-decreasing in $i$, then the optimal policy is a threshold policy.
\end{proposition}

\begin{proof} 
We proceed in two steps:

\medskip

\noindent \emph{Step 1:} For now, assume that $a(i,s)$ is \emph{generic}. That is, in equation \eqref{eq:dual_dp} we have  $a(i,s) -  \sum_{j>i} \sum_{s'}  u(j,s') \cdot  c(j,s',i,s) \neq 0$. In that case for each pair $(i,s)$ we have that either (a) $u(i,s) = 0$ and the dual constraint is non-tight; or (b) $u(i,s) > 0$ and the dual constraint is tight. Define the threshold $t(s)$ to be the smallest $i$ such that $u(i,s) > 0$. Then, because $u(i,s)$ is non-decreasing in $i$, for every $i < t(s)$ we are in case (a) and for every $i \geq t(s)$ we are in case (b).

By complementary slackness we have that in case (a) since the dual constraint is not tight, we must have $z(i,s) = 0$ and hence the probability of picking $i$ given signal $s$ and that $i$ is the best so far should be $q(i,s) = 0$ (by equation \eqref{eq:from-lp-to-policy}). In case (b) since $u(i,s) > 0$ the primal constraint should be tight and hence the probability of picking $i$ should be $1$ (again by equation \eqref{eq:from-lp-to-policy}). This shows that the policy must be a threshold policy.

\medskip

\noindent \emph{Step 2}: The only missing part now is to argue what happens if $a(i,s)$ is not generic. (This will be a technical and not particularly algorithmic argument. The reader that doesn't care about corner cases may want to skip it). In that case we can appeal to a perturbation argument: consider $a_\epsilon(i,s) = a(i,s) + \epsilon \cdot n(i,s)$ where $n(i,s)$ is a random perturbation and $\epsilon$ is a small number (that we will send to zero). Then almost surely we will have that the LP will be generic (in the sense of Step 2). Hence the optimal solution $z_\epsilon$ will be a threshold policy. Take now a sequence of $\epsilon_t \rightarrow 0$ and consider the threshold policy solutions $z_{\epsilon_t}$. Since the solutions $z_{\epsilon_t}$ live in a compact space, they must converge to some feasible solution $z^*$ in the limit (passing to a subsequence if necessary). This solution must also be a threshold policy since the set of threshold policies is closed. To see that this is an optimal solution to the unperturbed LP, note that the perturbation only affects the objective function, hence $z^*$ is feasible. Also note that for any feasible point $z$ we have that $\langle a_\epsilon, z_\epsilon \rangle \geq \langle a_\epsilon, z \rangle$. Taking the limit as $\epsilon \rightarrow 0$ we get: $\langle a, z^* \rangle \geq \langle a, z \rangle$ for all feasible $z$, hence $z^*$ is a solution to the unperturbed LP.
\end{proof}

For the reverse direction we have:

\begin{proposition}\label{prop:thresholds_greedy}
Assume NF and HI. If the optimal policy is a threshold policy, then the greedy backwards induction solution (equation \eqref{eq:dual_dp}) is optimal for the dual LP.
\end{proposition}
\begin{proof}
Let $t^*(s)$ be the thresholds in the optimal policy and let $u$ be any solution to the dual LP. Then the optimal primal solution is such that $z(i,s) = 0$ for $i < t^*(s)$ and the primal $(i,s)$-constraint is tight for $i \geq t^*(s)$.

This means in particular that for $i < t^*(s)$ the primal $(i,s)$ constraint is slack, so by complementary slackness we must have $u(i,s) = 0$. For $i \geq t^*(s)$ we have $z(i,s) > 0$ which means that the $(i,s)$-constraint needs to be tight (again by complementary slackness). It means in particular that for $i \geq t(s)$ we must have $u(i,s) = a(i,s) - \sum_{j>i}\sum_{s'} u(j,s) \cdot c(j,s',i,s)$.

Now we argue that $u(i,s)$ must be equal to the solution obtained by backwards induction in equation \eqref{eq:dual_dp}. If not, let $(i,s)$ be index with largest $i$ such they differ. Note that either the dual solution $u$ and the optimal solution are either zero or tight in each coordinate. If $a(i,s) - \sum_{j>i}\sum_{s'} u(j,s) \cdot c(j,s',i,s) > 0$ then neither can be tight so both must be zero. If $a(i,s) - \sum_{j>i}\sum_{s'} u(j,s) \cdot c(j,s',i,s) \leq 0$ then both need to be zero as well. So they must also agree on $(i,s)$.
\end{proof}

An example of a setting that satisfies HI and NF, but where the optimal policy is not a threshold policy is the following:

%\renato{ Can we get a counterexample with Esfandiari's model ?} 

\begin{example} Consider a Markovian stopping problem with $n=4$ transitions given by the chain in Figure \ref{fig:markovian_example}. The first state is $s_1 = 1$. The optimal policy is to stop at the second element if $s_2 = 2$ and otherwise to wait until the last element. This policy is memoryless but not a threshold policy since we stop at $s=2$ at $i=2$ but don't stop at the same signal at $i=3$.

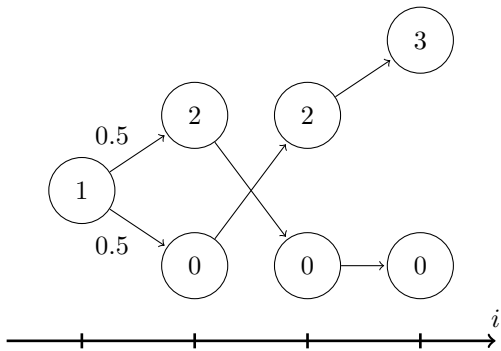
\begin{figure}[h!]
\centering
\usetikzlibrary{automata,positioning}
\begin{tikzpicture}
    \node[state] at (0,1) (1a) {$1$};
    \node[state] at (1.5,2) (2b) {$2$};
    \node[state] at (1.5,0) (0b) {$0$};
    \node[state] at (3,2) (2c) {$2$};
    \node[state] at (3,0) (0c) {$0$};
    \node[state] at (4.5,3) (3d) {$3$};
    \node[state] at (4.5,0) (0d) {$0$};
    \draw[every loop]
    (1a) edge node[auto=left] {$0.5$} (2b)
    (1a) edge node[auto=right] {$0.5$} (0b)
    (2b) edge node[auto=left] {} (0c)
    (0b) edge node[auto=left] {} (2c)
    (0c) edge node[auto=left] {} (0d)
    (2c) edge node[auto=left] {} (3d);
    \draw[->, line width=1] (-1,-1)--(5.5,-1);
    \draw[line width=1] (0,-1.1)--(0,-.9);
    \draw[line width=1] (1.5,-1.1)--(1.5,-.9);
    \draw[line width=1] (3,-1.1)--(3,-.9);
    \draw[line width=1] (4.5,-1.1)--(4.5,-.9);
    
    \node at (5.5,-.7) {$i$};
\end{tikzpicture}
\caption{Markovian stopping problem where the optimal policy is not a threshold policy}
\label{fig:markovian_example}
\end{figure}

In fact, in this example, the greedy algorithm yields an optimal solution to the dual LP (it sets $u(2,2) = 1/2$ and $u(4,3) = 1/2$ and so its objective value is $1$), but it violates the monotonicity assumption that is required for Proposition~\ref{prop:greedy_thresholds} to hold.
\end{example}

\section{Application: Secretary with Samples}\label{sec:samples}

As our first case study we consider the secretary problem with samples. For concreteness, we focus on the model of Kaplan et al.~\cite{DBLP:conf/soda/KaplanNR20} (Example~\ref{ex:samples}). Recall that in this model an adversary picks $n+k$ numbers. Then $k$ of these numbers chosen uniformly at random are shown to the algorithm as samples. Afterwards, the remaining $n$ values are presented in random order.

\paragraph{LP Formulation}

As signals we consider: $s =$ ``how many elements in the sample set are smaller than the current element''. So the signal space is in $\mathcal{S} = \{0, 1, 2, ... k\}$.

Now we need to calculate $a(i,s)$ and $c(i,s,j,s')$ and argue that both NF and HI are satisfied. The way to do that is based on the following way to sample:

\begin{itemize}
    \item Start with a total order on $k$ elements. Let's imagine that those are blue.
    \item Now we will add $n$ black elements one by one in the following way:\\
    For $t = 1, \dots, n$, there are $k+t$ positions where we can insert the $t$-th first black element. Insert in a random position.
\end{itemize}

This is the same sampling process as sampling the entire permutation and marking $k$ of those blue and revealing the other ones in random order. The sampling procedure decides $\preceq_{1..j}, s_1, \hdots, s_j$ before deciding the position of later elements $i > j$ implying NF. Similarly, since the remaining elements are inserted after $j$, the relative order and signals of elements before $j$ is irrelevant to whether $j$ will be the top element conditioned on $T_j,s_j$, therefore verifying HI. 
Now, from this sampling process we can compute the coefficients:

\begin{align*}
    a(i,s) &= \P(r_i= n, s_i=s)
    = \P(s_i = s) \cdot \P(r_i = n \mid s_i = s)
    = \frac{1}{k+1} \cdot \prod_{t=1}^{n-1} \frac{s+t}{k+t+1}.
\end{align*}

To see that imagine inserting first element $i$ (it must be inserted in one of the $k+1$ positions) and then inserting every other element and see what positions are available so that it is still the top so far.

For $c(i,s,j,s') = \P(T_j, s_j = s' \;|\; T_i, s_i = s)$ where $T_i$ means that $i$ is the top so far. Note that this is only non-zero if $j \leq i$ and $s' \leq s$. Subject to that we can again imagine inserting the elements one by one. We get:
\begin{align*}
c(i,s,j,s') = \frac{1}{s+1} \cdot \prod_{t=1}^{j-1} \frac{s'+t}{s+t+1}
\end{align*}
The idea is again very similar. Start by inserting $j$ and then insert all the elements between $1$ and $j-1$.

Plugging these formulas for the coefficients $a(i,s)$ and $c(i,s,j,s')$ into our generic LP, we obtain the following primal-dual pair:
$$
\begin{aligned}
&\max \sum_i \sum_s z(i,s) \cdot \frac{1}{k+1} \cdot \prod_{t=1}^{n-1} \frac{s+t}{k+t+1} \text{ s.t. } \\
& z(i,s) \leq 1 - \sum_{j<i} \sum_{s' \leq s} z(j,s') \cdot \frac{1}{s+1} \cdot \prod_{t=1}^{j-1} \frac{s'+t}{s+t+1}\\
& z(i,s) \geq 0
\end{aligned}
\qquad \left| \qquad
\begin{aligned}
&\min \sum_i \sum_s u(i,s) \text{ s.t. } \\
& u(i,s) + \sum_{j>i} \sum_{s' \geq s}  u(j,s') \cdot \frac{1}{s'+1} \cdot \prod_{t=1}^{i-1} \frac{s+t}{s'+t+1}\\
&\hspace*{40pt}\geq \frac{1}{k+1} \cdot \prod_{t=1}^{n-1} \frac{s+t}{k+t+1}\\
& u(i,s) \geq 0
\end{aligned}
\right.$$

See Figure~\ref{fig:samples} for plots of the success probability of the optimal policy for $n = 20$ and varying $k$ once as a function of $k$ and once as a function of $p = k/(k+n)$. 

\paragraph{Optimality of Thresholding}

Before we show optimality of thresholding, we establish the following lemma that will allow us to work with expressions arising in the dual (which can be thought as a form of discrete integration):

\begin{lemma}\label{lem:discrete_integral}
Given integers $a < b$ and $u<v$ the following equality holds:
$$\sum_{s=u}^v \prod_{t=a+1}^b (s+t) = \frac{1}{b-a+1} \left[ \prod_{t=a}^b (v+1+t) - \prod_{t=a}^b (u+t) \right]$$
\end{lemma}

\begin{proof} For some integer $s$ we have:
$$ \prod_{t=a}^b (s+1+t) - \prod_{t=a}^b (s+t) = \left[ (s+b+1)-(s+a) \right] \prod_{t=a+1}^b (s+t)  =   (b-a+1) \prod_{t=a+1}^b (s+t) $$
The result is then obtained by telescoping the above equality for $s=u$ to $v$.
\end{proof}

We can now prove:

\begin{theorem}\label{thm:samples_thresholds}
For every $n$ and every $k$ the optimal policy for secretaries with samples is a threshold policy.
\end{theorem}

\begin{proof}[Proof Sketch (Full Proof in Appendix~\ref{apx:samples})]

We prove the theorem through proposition Proposition~\ref{prop:greedy_thresholds}. We need to show that  the greedy solution in equation \eqref{eq:dual_dp} is optimal and non-decreasing in $i$. Monotonicity follows from observing that  $a(i,s) = \frac{1}{k+1} \cdot \prod_{t=1}^{n-1} \frac{s+t}{k+t+1}$ does not depend on $i$ and $c(j,s',i,s) = \frac{1}{s'+1} \cdot \prod_{t=1}^{i-1} \frac{s+t}{s'+t+1}$ is non-increasing in $i$.\\

For optimality assume that the optimal solution $u(i,s)$ does not satisfy the recursion that defines the greedy algorithm. Then there must be a largest index $i^*$ and a signal $s^*$ for which the equation is violated. Now we consider changing $u(i,s)$ to $u(i,s) + \Delta(i,s)$ where $\Delta(i^*,s^*) = - \delta$ for
\[
    \delta := u(i^*,s^*) - \max\left\{0,a(i^*,s^*) - \sum_{j > i^*} \sum_{s' \geq s} u(j,s') \cdot c(j,s',i^*,s^*) \right\}
\]
and $\Delta(i,s) = 0$ for all other $i \geq i^*$. For $i < i^*$ we define $\Delta(i,s)$ recursively as follows
\[
    \Delta(i,s) = - \sum_{j=i+1}^{i^*} \sum_{s' \geq s} \Delta(j,s') \cdot c(j,s',i,s).
\]
Furthermore, define
\[
    R_i := \sum_{j \geq i} \sum_{s \leq s^*} \Delta(j,s)
\]
and note that $R_i$ is the cumulative change to the objective function. We can now prove by backward induction that (see full proof in Appendix \ref{apx:samples} for the details):

\begin{align*}
%% first property
\text{(i)~ For $i = i^*$:}\quad
&\qquad\Delta(i^*,s^*) = - \delta \quad\text{for $s = s^*$, and}\\ &\qquad\Delta(i^*,s) = 0 \quad\text{for all $s \neq s^*$}.\\[4pt]  
\text{\phantom{(i)~} For $i < i^*$:}\quad 
&\qquad\Delta(i,s) =  
\delta \cdot \frac{1}{s^*+1} \prod_{t=1}^{i^*-2} \frac{s+t}{s^*+t+1}
\quad \text{for all $s \leq s^*$, and}\\[6pt]
&\qquad \Delta(i,s) = 0 \quad \text{for all $s > s^*$}.\\
%% second property
\text{(ii)}\quad \text{For all $i \leq i^*$:} 
&\qquad R_i = - \delta \cdot \frac{i-1}{i^*-1} \leq 0.
\end{align*}

In fact, by the definition of $R_i$ and (i), we have
\begin{align*}
R_i 
= \left(\sum_{j = i}^{i^*-1} \sum_{s = 0}^{s^*} \Delta(j,s) \right) - \delta 
&= \delta \cdot \left( \sum_{j = i}^{i^*-1} \sum_{s=0}^{s^*}  \frac{1}{s^*+1} \prod_{t=1}^{i^*-2} \frac{s+t}{s^*+t+1} \right) - \delta \\
&= \delta \cdot \sum_{j=i}^{i^*-1} \left( \frac{
\sum_{s=0}^{s^*} \prod_{t=1}^{i^*-2} (s+t)
}{
\prod_{t=1}^{i^*-1} (s^*+t)
}\right) - \delta 
= \left(\frac{i^*-i}{i^*-1}-1\right) \delta = - \delta \cdot \frac{i-1}{i^*-1} \leq 0,
\end{align*}
where the last equality follows from Lemma~\ref{lem:discrete_integral}.

We now use (i) and (ii) to argue that the operation preserves feasibility and only improves the objective: 
We have chosen the recursion for the $\Delta(i,s)$ to satisfy the first constraint of the dual. From (i) we get that all the $\Delta(i,s)$ are non-negative, so we also satisfy the non-negativity constraints. From (ii) we get that $R_1 \leq 0$ so we only decrease the objective.

By repeatedly applying this operation we can conclude that the greedy solution is an optimal solution, just as we claimed.
\end{proof}

An immediate implication of Theorem~\ref{thm:samples_thresholds} is an efficient (poly-time) algorithm for computing the optimal policy, and the winning probability of that policy.

\paragraph{Explicit solution to dual}

Next we derive an explicit (non-recursive) formula for the optimal dual solution.

\begin{theorem}
\label{thm:samples_explicit_dual}
The following is an explicit solution to the dual recursion:

For $i = n$ and all $s$:
\begin{align*}
u(n,s) = \frac{1}{k+1} \cdot \prod_{t=1}^{n-1} \frac{s+t}{k+t+1}.
\end{align*}

For $i < n$ and all $s$:
\begin{align*}
    u(i,s) &= \max\Bigg\{0, \frac{1}{k+1} \cdot \prod_{t=1}^{i-1} \bigg(\frac{s+t}{k+t+1}\bigg) \cdot \Bigg( \prod_{t=i}^{n-1} \bigg(\frac{s+t}{k+t+1}\bigg) -\\
    &\hspace*{170pt}\frac{\sum_{j=1}^{n-i} \left({n-i \choose j} \cdot \frac{1}{j} \cdot \prod_{\ell=1}^{j} (k-s+\ell) \cdot \prod_{\ell=1}^{n-i-j} (s+i-1+\ell)\right)}{\prod_{t=i}^{n-1} (k+t+1)} \Bigg)\Bigg\},
\end{align*}
where we use the convention that $\prod_{t=a}^{b} x_t = 1$ if $b < a$.
\end{theorem}

We note that the dual solution has the following natural interpretation: The positive term is the probability that secretary $i$ with signal $s$ is the best over all. This is the winning probability if we would accept. The negative terms are the probability with which we would win if we would pick the first secretary among the remaining $n-i$ secretaries that is better than the current one. 

\begin{proof}[Proof of Theorem~\ref{thm:samples_explicit_dual}]
For the proof we can ignore the $\max\{0,\dot\}$.
We prove the claim by induction. The base case ($i = n$ and all $s$) holds by definition. Now let's do the inductive step. Assume the claim holds for all $i' > i$ and all $s$. Then we can use the induction hypothesis to obtain
\[
u(i,s) = \frac{1}{k+1} \cdot \prod_{t=1}^{n-1} \left(\frac{s+t}{k+t+1}\right) - \frac{1}{k+1} \cdot  \prod_{t=1}^{n-1} \left( \frac{1}{k+t+1} \right) \cdot \prod_{t=1}^{i-1} (s+t) \cdot \sum_{s' = s}^{k} T(i,s'), 
\]
where 
\begin{align*}
T(i,s') 
&= (n-i) \prod_{t=i+1}^{n-1} (s'+t) - \sum_{z = i+1}^{n-1} \left( \sum_{j=1}^{n-z} \left({n-z \choose j} \cdot \frac{1}{j} \cdot \prod_{\ell = 1}^{j} (k-s'+\ell) \cdot \prod_{t=i+1}^{n-j-1} (s'+t) \right)\right) \\
&= (n-i) \prod_{t=i+1}^{n-1} (s'+t) - \sum_{j=1}^{n-i-1} \left(\sum_{z = i+1}^{n-j} \left( {n-z \choose j} \cdot \frac{1}{j}\right) \cdot \prod_{\ell = 1}^{j} (k-s'+\ell) \cdot \prod_{t=i+1}^{n-j-1} (s'+t) \right)\\
&= (n-i) \prod_{t=i+1}^{n-1} (s'+t) - \sum_{j = 1}^{n-i-1} \left( (n-i) \cdot {n-i-1 \choose j} \cdot \frac{1}{j(j+1)} \cdot \prod_{\ell = 1}^{j} (k-s'+\ell) \cdot \prod_{t=i+1}^{n-j-1} (s'+t)\right).
\end{align*}

In particular, to establish the claim it suffices to show that
\[
\sum_{s' = s}^{k} T(i,s') = \sum_{j=1}^{n-i}\left( {n-i \choose j} \cdot \frac{1}{j} \cdot \prod_{\ell =1}^{j} (k-s+\ell) \cdot \prod_{t=i}^{n-j-1} (s+t)\right). 
\]

We prove this identity by backward induction over $s$. For the base $s = k$ case we need to show
\[
T(i,k) = \sum_{j=1}^{n-i} \left({n-i \choose j} \cdot \frac{1}{j} \cdot j! \cdot \prod_{t = i}^{n-j-1} (k+t)\right).
\]

Indeed, we have
\begin{align*}
    %% first line
    &\sum_{j=1}^{n-i} \left({n-i \choose j} \cdot \frac{1}{j} \cdot j! \cdot \prod_{t=i}^{n-j-1} (k+t) \right) \\
    %% second line
    %&\qquad= \sum_{j=1}^{n-i-1} \left({n-i \choose j} \cdot \frac{1}{j} \cdot j! \cdot \prod_{t=i}^{n-j-1} (k+t) \right) + (n-i-1)!\\
    %% third line
    %&\qquad= \sum_{j=1}^{n-i-1} \left({n-i \choose j} \cdot \frac{1}{j} \cdot j! \cdot (k+i) \cdot \prod_{t=i+1}^{n-j-1} (k+t) \right) + (n-i-1)!\\
    %% fourth line
    &\qquad= \sum_{j=1}^{n-i-1} \left({n-i \choose j} \cdot \frac{1}{j} \cdot j! \cdot \bigg((k+n-j)-(n-i-j)\bigg) \cdot \prod_{t=i+1}^{n-j-1} (k+t) \right) + (n-i-1)! \\
    %% fifth line
    &\qquad= (n-i) \cdot \prod_{t=i+1}^{n-1} (k+t) + \sum_{j=2}^{n-i-1} \left({n-i \choose j} \cdot \frac{1}{j} \cdot j! \cdot (k+n-j) \cdot  \prod_{t=i+1}^{n-j-1} (k+t)\right) + (n-i-1)!\\
    %% sixth line
    &\hspace*{122pt}- \sum_{j=1}^{n-i-1} \left({n-i \choose j} \cdot \frac{1}{j} \cdot j! \cdot (n-i-j) \cdot  \prod_{t=i+1}^{n-j-1} (k+t)\right) \\
    %% seventh line
    &\qquad= (n-i) \cdot \prod_{t=i+1}^{n-1} (k+t) + \sum_{j=1}^{n-i-2} \left({n-i \choose j+1} \cdot \frac{1}{j+1} \cdot (j+1)! \cdot (k+n-j-1) \cdot  \prod_{t=i+1}^{n-j-2} (k+t)\right) + (n-i-1)!\\
    %% eighth line
    &\hspace*{122pt}- \sum_{j=1}^{n-i-1} \left({n-i \choose j}  \cdot  \frac{1}{j} \cdot j! \cdot (n-i-j) \cdot  \prod_{t=i+1}^{n-j-1} (k+t)\right) \\
    %% nineth line
    &\qquad= (n-i) \cdot \prod_{t=i+1}^{n-1} (k+t) + \sum_{j=1}^{n-i-2} \left({n-i \choose j+1} \cdot \frac{1}{j+1} \cdot (j+1)! \cdot  \prod_{t=i+1}^{n-j-1} (k+t)\right) + (n-i-1)!\\
    %% tenth line
    &\hspace*{122pt}- \sum_{j=1}^{n-i-1} \left({n-i \choose j} \cdot \frac{1}{j} \cdot j! \cdot (n-i-j) \cdot  \prod_{t=i+1}^{n-j-1} (k+t)\right) \\
    %% eleventh line
    &\qquad= (n-i) \cdot \prod_{t=i+1}^{n-1} (k+t) + \sum_{j=1}^{n-i-1} \left({n-i \choose j+1} \cdot  \frac{1}{j+1} \cdot (j+1)! \cdot  \prod_{t=i+1}^{n-j-1} (k+t)\right)\\
    %% twelfth line
    &\hspace*{122pt}- \sum_{j=1}^{n-i-1} \left({n-i \choose j} \cdot \frac{1}{j} \cdot j! \cdot (n-i-j) \cdot  \prod_{t=i+1}^{n-j-1} (k+t)\right) \\
    %% thirteenth line
    &\qquad= (n-i) \cdot \prod_{t=i+1}^{n-1} (k+t) - \sum_{j=1}^{n-i-1} \left(\left({n-i \choose j} \cdot \frac{1}{j} \cdot j! \cdot (n-i-j) - {n-i \choose j+1} \frac{1}{j+1} \cdot (j+1)!\right) \prod_{t=i+1}^{n-j-1} (k+t) \right)\\
    %% fourteenth line
    &\qquad = (n-i) \cdot  \prod_{t=i+1}^{n-1} (k + t) - \sum_{j=1}^{n-i-1} \left( (n-i) \cdot {n-i-1 \choose j} \cdot \frac{1}{j(j+1)} \cdot j! \cdot  \prod_{t= i+1}^{n-j-1} (k+t) \right) = T(i,k),
\end{align*}
as claimed.

The argument for the inductive step is similar, but technically a bit more involved. We defer the details to Appendix~\ref{apx:inductive_step}.
\end{proof}

\paragraph{Gilbert and Mosteller as the limit when $\mathbf{k \rightarrow \infty}$}

In the model of Gilbert and Mosteller (Example \ref{ex:gm}) each secretary is associated with a sample from a known distribution $F$. The distribution can without loss of generality be thought as the uniform distribution over $[0,1]$ since the algorithm can always process the quantiles $q_i = F^{-1}(s_i) \sim U[0,1]$. Quantiles can be seen as a limit of the secretary with samples model with $k \rightarrow \infty$ by taking  $q = s/(k+1)$ where $s \in \{0, \dots, k+1\}$ and take the limit $k \rightarrow \infty$. In the limit, $q \in [0,1]$ corresponds to the quantile of the secretary.
One can then take the limit of the LP coefficients:
\begin{align*}
    a(i,q) &= q^{n-1} \;dq \quad \text{and} \quad
     c(i,q,j,q') = \frac{1}{q} \left(\frac{q'}{q}\right)^{j-1} \;dq'
\end{align*}
and obtain the following functional optimization problem in the limit:
$$\begin{aligned}
& \max \sum_{i} \int_s z(i,q) \cdot q^{n-1} \;dq \text{ s.t. } \\
& z(i,q) \leq 1 - \sum_{j<i} \int_{q' \leq q} z(j,q') \cdot \frac{1}{q} \cdot \left(\frac{q'}{q}\right)^{j-1} \;dq'\\
& z(i,q) \geq 0
\end{aligned} \qquad \left| \qquad \begin{aligned}
& \min \sum_i \int_s u(i,q) \; dq \text{ s.t. } \\
& u(i,q) + \sum_{j>i} \int_{q' \geq q} u(j,q') \cdot \frac{1}{q'} \cdot \left(\frac{q}{q'}\right)^{i-1} \;dq' \geq q^{n-1} \\
& u(i,q) \geq 0
\end{aligned} \right.$$
Since this is a limit of secretaries with samples, we can again obtain a dual solution via backwards induction, which has a particularly nice form:
\begin{corollary}[From Theorem~\ref{thm:samples_thresholds}]\label{cor:explicit-dual-solution} 
The optimal solution $u(i,s)$ to the dual LP found by the greedy backward induction algorithm satisfies
\begin{align*}
{u(i,s)} & = s^{n-1} &&\text{{for $i = n$ and all $s$, and}}\\
{u(i,s)} &
= \max\left\{0, s^{i-1}\cdot\left(s^{n-i} - \sum_{k=1}^{n-i} \frac{1}{k} {n-i \choose k} s^{n-i-k} (1-s)^k\right)\right\} &&\text{{for $i < n$ and all $s$.}}
\end{align*}
\end{corollary}
An immediate consequence of Corollary~\ref{cor:explicit-dual-solution} is that the optimal policy can be determined by setting $s^*_n = 0$ and finding for each $i< n$ the $s^*_i$ such that
\begin{align}
{(s^*_i)^{n-i} - \sum_{k=1}^{n-i} \frac{1}{k} {n-i \choose k} (s^*_i)^{n-i-k} (1-(s^*_i))^k = 0,} \label{eq:gm-decision-numbers}
\end{align}
and to then accept secretary $i$ with signal $s$ if it is the best so far and $s \geq s^*_i$. We note that this is precisely how Gilbert and Mosteller \cite{gilbert2006recognizing} define the optimal policy (Equation (3b-1) on p.~53).

\paragraph{Winning Probability}

More generally, the explicit dual solution (in Theorem \ref{thm:samples_explicit_dual} and Corollary~\ref{cor:explicit-dual-solution}) allows to deduce thresholds in signal space (as we just did for Gilbert and Mosteller), both in the non-asymptotic and asymptotic regimes. For Gilbert and Mosteller, for example, it is known that solutions to equation~(\ref{eq:gm-decision-numbers}) in the asymptotic regime satisfy $b_i \rightarrow 1$ and $i(1-b_i) \rightarrow c$, where $b_i$ for $i = n-1, \dots, 0$ is the threshold when there are $i$ secretaries after the current one, and this can be used to obtain analytic expressions for the asymptotic winning probability \cite{gnedin96,Samuelson82}. 

%% SECTION 6
\section{Application: Advice from a Binary Classifier}\label{sec:binary}

We will now use the technology developed in the previous sections to derive the optimal policy for the secretary problem with advice from a binary classifier (from the intro, and defined formally in Section~\ref{sec:moodel}). Recall that in this example secretaries have uniform random ranks, and that for the top secretary (rank = n) we receive advice $Y$ with probability $p$ and advice $N$ with probability $1-p$. For all other secretaries we receive advice $N$ with probability $p'$ and advice $Y$ with probability $1-p'$. In ML speak, $p$ corresponds to the precision of the advice and $p'$ to its specificity.

\paragraph{Notation} To avoid polluting the expressions with too many parenthesis, in this section we will abbreviate the primal $z(i,s)$ and the dual $u(i,s)$ using $z_i^s$ and $u_i^s$. Moreover, we will use the notation:
$$\bar z_j = (1-p') z_j^Y + p' z_j^N.$$
%Finally we will abbreviate the thresholds $t(Y)$ and $t(N)$ by $t_Y$ and $t_N$.\\

\paragraph{LP Formulation} Since Example~\ref{ex:yes_no} satisfies NF and HI we can instantiate our general linear programming formulation to the binary classifier case. Using the notation established above, we can write for $n$ periods and parameters $p$ and $p'$:
$$\begin{aligned}
& \max \sum_i \frac{1}{n}\left[p z_i^Y + (1-p) z_i^N\right] \text{ s.t. } \\
& z_i^s \leq 1 - \sum_{j<i} \frac{1}{j}\left[(1-p') z_j^Y + p' z_j^N\right]\\
& z_i^s \in [0,1]
\end{aligned} \qquad \left| \qquad \begin{aligned}
& \min \sum_i \left(u_i^Y + u_i^N\right) \text{ s.t. } \\
& u_i^Y + \sum_{j>i} \frac{1-p'}{i} \cdot \left( 
u_j^Y + u_j^N 
\right) \geq \frac{p}{n}\\
& u_i^N + \sum_{j>i} \frac{p'}{i} \cdot \left( 
u_j^Y + u_j^N 
\right) \geq \frac{1-p}{n}\\
& u_i^Y, u_i^N \geq 0
\end{aligned} \right.$$

\paragraph{Optimality of Thresholding} 
We start by using the dual LP to show:

\begin{theorem}\label{thm:binary_thresholds}
The optimal policy for the binary advice problem is a threshold policy.
\end{theorem}

\begin{proof}[Proof Sketch (Full Proof in Appendix~\ref{apx:binary})] The proof uses Proposition~\ref{prop:greedy_thresholds}, by which it suffices to shows that the greedy algorithm yields an optimal and monotone solution to the dual LP. Our argument follows the same blueprint as the proof of  Theorem~\ref{thm:samples_thresholds}. Monotonicity follows from the properties of the coefficients $a(i,s)$ and $c(j,s',i,s)$. For optimality we use the same basic construction, with appropriately adjusted inductive claims.
\end{proof}

\paragraph{The Optimal Threshold Policy}

Next we use the fact that the optimal policy is a threshold policy to obtain a closed-form understanding of the optimal policy and the performance it achieves. 

Let $t_N$ and $t_Y$ denote the thresholds (in time) after which we start accepting the best-ranking secretary so far conditioned on the advice being No ($N$) or Yes ($Y$).
Write $\textsf{Opt}(n,p,p')$ for the optimal solution to the primal LP, and let $\textsf{OPT}(p,p') = \inf_{n \geq 1} \textsf{Opt}(n,p,p')$.

\begin{theorem}\label{thm:binary_advice}
With advice from a binary classifier with recall $p$ and specificity $p'$, the optimal policy is a threshold policy that has two thresholds $t_Y \leq t_N$ from which onwards it accepts the best secretary so far upon receiving advice $Y$ or $N$, and the optimal objective is: 
\[
\textsf{Opt}(p,p') \geq \lim_{n \rightarrow \infty} \textsf{Opt}(n,p,p') =  \left(\frac{1-p'}{p}\right)^{(1-p')/p} \cdot e^{p/p'-1/p'}.
\]
The two thresholds are $t_Y = n \cdot (\frac{1-p'}{p})^{1/p'}\cdot e^{p/p'-1/p'}$ and $t_N = n \cdot e^{p/p'-1/p'}$.
\end{theorem}

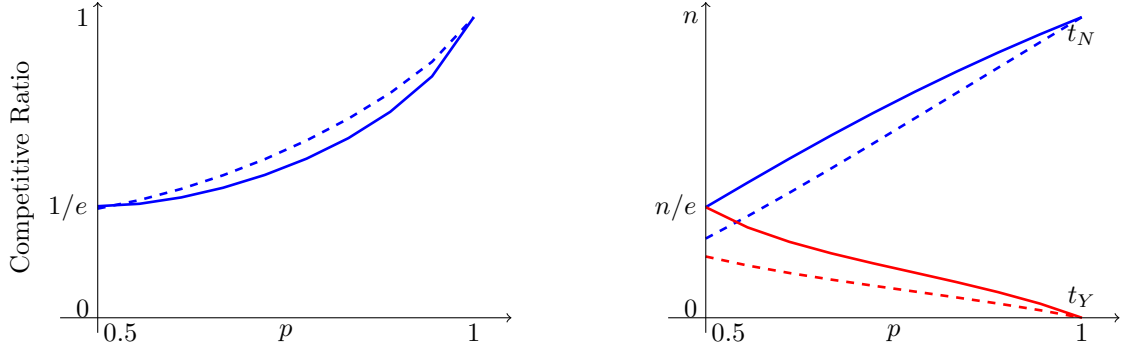
\begin{figure}[t]
\centering
\begin{tikzpicture}[xscale=10, yscale=4]
\draw[line width=1,  color=blue] (0.5, 0.37104277871264424)--
 (0.55555555555555558, 0.37910695541424322)--
 (0.61111111111111116, 0.4003136327102359)--
 (0.66666666666666663, 0.43244764221685245)--
 (0.72222222222222221, 0.47513807686812032)--
 (0.77777777777777779, 0.52935023651537838)--
 (0.83333333333333326, 0.59758322965465616)--
 (0.88888888888888884, 0.68502240756169819)--
 (0.94444444444444442, 0.80291350237634862)--
 (1.0, 0.99999999999998879);

\draw[dashed, line width=1, color=blue] (0.5, 0.3630407264452067) -- (0.5555555555555556, 0.3914314541059371) -- (0.6111111111111112, 0.4286481330300725) -- (0.6666666666666666, 0.47398785011707917) -- (0.7222222222222222, 0.5274873574433466) -- (0.7777777777777778, 0.5897909280316999) -- (0.8333333333333333, 0.6622734836108766) -- (0.8888888888888888, 0.7475651248527397) -- (0.9444444444444444, 0.8514118195804796) -- (1.0, 1.0);
 
\draw[->] (0.45,0) -- (1.05,0);
\draw[->] (0.5,-0.05) -- (0.5,1.05);
\node at (.75,-.05) {$p$};
\node at (.53,-.05) {$0.5$};
\node at (1,-.05) {$1$};
\node at (.46,.367) {$1/e$};
\node at (.48,.03) {$0$};
\node at (.48,1) {$1$};
\node[rotate=90] at (.4,.5) {$\text{Competitive Ratio}$};

\begin{scope}[xshift=23]
\draw[line width=1, color=blue] (0.5, 0.36787944117144233) -- (0.55555555555555558, 0.4493289641172217) -- (0.61111111111111116, 0.5292133415000504) -- (0.66666666666666663, 0.6065306597126334) -- (0.72222222222222221, 0.6807123983233854) -- (0.77777777777777779, 0.751477293075286) -- (0.83333333333333326, 0.8187307530779817) -- (0.88888888888888884, 0.8824969025845955) -- (0.94444444444444442, 0.9428731438548749) -- (1.0, 1.0);

\draw[dashed, line width=1,  color=blue] (0.5, 0.2635971381157267) -- (0.5555555555555556, 0.33766051365513206) -- (0.6111111111111112, 0.41572015612511054) -- (0.6666666666666666, 0.4965853037914095) -- (0.7222222222222222, 0.5793168542160426) -- (0.7777777777777778, 0.6631763835015962) -- (0.8333333333333333, 0.7475836370778212) -- (0.8888888888888888, 0.8320826289821267) -- (0.9444444444444444, 0.9163151181838687) -- (1.0, 1.0);

%\draw[line width=1,  color=green] 
%(0.5, 0.2635971381157267) -- (0.5555555555555556, 0.33766051365513206) -- (0.6111111111111112, 0.41572015612511054) -- (0.6666666666666666, 0.4965853037914095) -- (0.7222222222222222, 0.5793168542160426) -- (0.7777777777777778, 0.6631763835015962) -- (0.8333333333333333, 0.7475836370778212) -- (0.8888888888888888, 0.8320826289821267) -- (0.9444444444444444, 0.9163151181838687) -- (1.0, 1.0);
\draw[dashed, line width = 1, color = red] (0.5, 0.20374971871270472) -- (0.5555555555555556, 0.17385795647277472) -- (0.6111111111111112, 0.14892653164319494) -- (0.6666666666666666, 0.1270018998300289) -- (0.7222222222222222, 0.10676213545097504) -- (0.7777777777777778, 0.08723506409707778) -- (0.8333333333333333, 0.06762826971116472) -- (0.8888888888888888, 0.047201517555273124) -- (0.9444444444444444, 0.025115840416584483) -- (1.0, 0.0);

\draw[line width=1, color = red]
(0.5, 0.36787944117144233) -- (0.55555555555555558, 0.30069512768373236) -- (0.61111111111111116, 0.25259303137532102) -- (0.66666666666666663, 0.21444097124017678) -- (0.72222222222222221, 0.18129472962122328) -- (0.77777777777777779, 0.15010660595931891) -- (0.83333333333333326, 0.11867987997168922) -- (0.88888888888888884, 0.085062267284683801) -- (0.94444444444444442, 0.046948854029558049) -- (1.0, 0.0);

\draw[->] (0.45,0) -- (1.05,0);
\draw[->] (0.5,-0.05) -- (0.5,1.05);
\node at (.75,-.05) {$p$};
\node at (.53,-.05) {$0.5$};
\node at (1,-.05) {$1$};
\node at (.46,.367) {$n/e$};
\node at (.48,.03) {$0$};
\node at (.48,1) {$n$};
\node at (1,.94) {$t_N$};
\node at (1,0.07) {$t_Y$};
\end{scope}
\end{tikzpicture}
\caption{Left plot shows the performance of the optimal secretary with advice policy. The right plot shows the thresholds $t_Y$ and $t_N$. In both cases, solid lines indicate $p = p$ and the dashed lines show $p' = (1 + p)/2$. As we see, higher accuracy (dashed lines) leads to increased competitive ratio.}
\label{fig:aproximation_plot_assym}
\end{figure}

Note how with $p = p' = 1/2$ (no advice), both thresholds become $t_Y = t_N = n/e$ recovering Dynkin's algorithm. If $p=p'=1$ (perfect advice) then the thresholds become $t_Y = 0$ and $t_N = n$ which means the policy should always pick whenever the signal is $Y$ and never pick if the signal is $N$. Figure \ref{fig:aproximation_plot_assym}  shows how the approximation and thresholds vary as a function of $p$ for the symmetric case $p=p'$, and asymmetric case $p' = (1 + p)/2$.

\begin{proof}[Proof of Theorem \ref{thm:binary_advice}]
Throughout the proof we will assume $n$ is large and we will approximate Riemann sums by integrals  $\sum_{i=1}^n \frac{1}{n} f(\frac{i}{n}) \approx \int_0^1 f(x) dx$. Whenever we use the symbol $\approx$ what we really mean is an equality up to terms that vanish in the limit as $n \rightarrow \infty$ which we choose to omit to prevent making the notation too ugly. As an example, whenever $a,b = \Theta(n)$ we will approximate the harmonic sums $\sum_{j=a}^b \frac{1}{j}$ by $\log(b/a)$.

\medskip

Theorem \ref{thm:binary_thresholds} gives a recipe on how to calculate the threshold using the dual LP.
Recall the dual LP for binary advice from the beginning of this section. For notational convenience we will work with the following variant, in which we moved the $1/n$ term from the constraints to the objective:
$$\min \sum \frac{u_i^Y}{n} + \frac{u_i^N}{n} \text{ s.t. } u_i^Y + \sum_{j > i} \frac{1-p'}{i} \left[u_j^Y + u_j^N\right] \geq p \text{ and } u_i^N + \sum_{j > i} \frac{p'}{i} \left[u_j^Y + u_j^N\right] \geq 1-p\text{ and } u_i^a \geq 0.$$

By the structure of the dual and a backward induction argument we have that for $i \geq t_N$ the dual has the following form:
$$u_i^Y = p - \sum_{j=i}^{n-1} \frac{1-p'}{j} \quad \text{and} \quad u_i^N = (1-p) - \sum_{j=i}^{n-1} \frac{p'}{j}.$$
The threshold $t_N$ is defined as the first time in which $u_i^N$ becomes zero. It is convenient at this point to approximate the harmonic sums in $u_i^N$ by a logarithm:
$$u_i^N \approx 1-p - p' \log\left( \frac{n}{i} \right),$$
which vanishes at $t_N = n \cdot e^{p/p'-1/p'}$. Now for $i < t_N$ we update only $u_i^Y$, then the formula becomes:
$$u_i^Y = p - \frac{1-p'}{i} \left[ u_{i+1}^Y + (p-u_{i+1}^Y ) \frac{i+1}{1-p'} \right] =  \frac{i+p'}{i}  u_{i+1}^Y - \frac{p}{i}.$$
Solving for the recursion we get the following:
$$u_i^Y = u_{t_N}^Y \cdot \prod_{j=i}^{t_N-1} \left( 1 + \frac{p'}{j}\right) - \sum_{j=i}^{t_N-1} \frac{p}{j}  \prod_{k=i}^{j-1} \left( 1 + \frac{p'}{k}\right).$$
Using $1+x \approx e^{-x}$ since $x = 1/j$ for $j = \Omega(n)$ and then approximating harmonic sums by logs, we get:
$$u_i^Y \approx 
u_{t_N}^Y \cdot e^{p' \log\left( t_N / i \right)} - \sum_{j=i}^{t_N-1} \frac{p}{j}  e^{p' \log(j/i)} = u_{t_N}^Y \cdot \left( \frac{t_N}{i} \right)^{p'} - \sum_{j=i}^{t_N-1} \frac{p}{j}  \left( \frac{j}{i} \right)^{p'}.   $$

For the last term we observe it can be interpreted as a Riemann sum and therefore approximated as the corresponding integral:
$$\sum_{j=i}^{t_N-1} \frac{p}{j}  \left( \frac{j}{i} \right)^{p'}\approx \frac{p}{i^{p'}} \int_i^{t_N} x^{p'-1} dx = \frac{p}{i^{p'}} \frac{1}{p'} [t_N^{p'} - i^{p'}] = \frac{p}{p'} \left[\left( \frac{t_N}{i} \right)^{p'} - 1\right].$$
%Using our approximation for $i\geq t_N$, $u_i^Y = p - (1-p') \log(n/i)$ and the value of $t_N = n \cdot e^{p/p'-1/p'}$ we get $u_{t_N}^Y = \frac{p}{p'}-\frac{1-p'}{p'}$ and 
Thus,
$$u_i^Y = \frac{p}{p'} + \left(u_{t_N}^Y - \frac{p}{p'}\right) \left(\frac{t_N}{i}\right)^{p'} = \frac{p}{p'} - \left(\frac{1-p'}{p'}\right) \left(\frac{t_N}{i}\right)^{p'}$$
vanishing when:
$$i = t_N \left(\frac{1-p'}{p}\right)^{1/p'} =: t_Y.$$

Now we can sum the dual to get the dual objective:
$$\begin{aligned}\textsc{DualObj} & \approx \frac{1}{n} \int_{t_Y}^{t_N} \left( \frac{p}{p'} - \frac{1-p'}{p'} \cdot \left( \frac{t_N}{x} \right)^{p'} \right) dx + \frac{1}{n}  \int_{t_N}^{n} \left( 1-\log\left( \frac{n}{x} \right) \right) dx \\ & =
\frac{t_N - t_Y}{n} \frac{p}{p'} - \frac{1}{n} \cdot \frac{1-p'}{p'} \cdot \frac{t_Y \left(\frac{t_N}{t_Y}\right)^{p'}-t_N}{p'-1} + \frac{t_N}{n} \log\left(\frac{n}{t_N}\right) \\
&= \frac{t_N - t_Y}{n} \frac{p}{p'} + \frac{t_N}{n} \log\left(\frac{n}{t_N}\right) - \frac{1}{np'} \left[t_N - t_N^{p'} t_Y^{1-p'} \right] \\ & = \frac{1}{np'} t_N^{p'} t_Y^{1-p'} - \frac{p}{p'} \frac{t_Y}{n}
\\&= \left(\frac{1-p'}{p}\right)^{(1-p')/p'} \cdot e^{p/p'-1/p'}.
\end{aligned}$$
This concludes the proof.
\end{proof}

%% SECTION 7
\section{Discussion and Future Work} 
\label{sec:discussion}
In this work we present a unifying view of multiple versions of the classic secretary problem. The linear program that is at the heart of our analysis precisely isolates the effect of the signalling scheme from the combinatorial structure of the problem: the joint distribution on ranks and signals only affects the coefficients of the simple set of constraints. The formulations that are captured by our analysis are quite diverse, as we demonstrate through the examples in Section \ref{sec:moodel}, and is it remarkable that all of these can be solved optimally by memoryless policies. 

As we saw in Example \ref{ex:no_nfhi}, however, some settings require simple, but not memoryless policies. One possible direction for future work is to explore {\em counting} policies, that do not rely on the permutation of signals observed, but only on their histograms. Another approach is to consider advice that does not satisfy NF or HI, but still has sufficient structure to reason about its efficacy.

More broadly, we hope that the advice lens can be used to abstract the modeling assumptions and find formal connections between related problems in other areas.

%% REFERENCES
\bibliographystyle{abbrv}
\bibliography{biblio}

%% APPENDIX
\appendix

\section{Full Proof of Theorem~\ref{thm:samples_thresholds}}\label{apx:samples}

\begin{proof}[Full Proof of Theorem \ref{thm:samples_thresholds}]

We prove the theorem through  Proposition~\ref{prop:greedy_thresholds}. We need to show that  the greedy solution obtained via backwards induction is optimal and non-decreasing in $i$. The greedy algorithms yields:

\begin{align*}
    &u(n,s) = \frac{1}{k+1} \cdot \prod_{t=1}^{n-1} \frac{s+t}{k+t+1} &&\text{and}\\
    &u(i,s) = \max\left\{0, \frac{1}{k+1}\cdot \prod_{t=1}^{n-1} \frac{s+t}{k+t+1} - \sum_{j>i} \sum_{s' \geq s}  u(j,s') \cdot \frac{1}{s'+1} \cdot \prod_{t=1}^{i-1} \frac{s+t}{s'+t+1}\right\} &&\text{for $i < n$ and all $s$.}
\end{align*}

Monotonicity follows from observing that  $a(i,s) = \frac{1}{k+1} \cdot \prod_{t=1}^{n-1} \frac{s+t}{k+t+1}$ does not depend on $i$ and $c(j,s',i,s) = \frac{1}{s'+1} \cdot \prod_{t=1}^{i-1} \frac{s+t}{s'+t+1}$ is non-increasing in $i$.\\

It remains to show optimality. 
Suppose the optimal dual solution does not satisfy the recursion that defines the greedy algorithm. Then there must be a largest index $i^*$ and a signal $s^*$ for which the equation is violated. Now we consider changing $u(i,s)$ to $u(i,s) + \Delta(i,s)$ where $\Delta(i^*,s^*) = - \delta$ for
\[
    \delta := u(i^*,s^*) - \max\left\{0,a(i^*,s^*) - \sum_{j > i^*} \sum_{s' \geq s} u(j,s') \cdot c(j,s',i^*,s^*) \right\}
\]
and $\Delta(i,s) = 0$ for all other $i \geq i^*$. For $i < i^*$ we define $\Delta(i,s)$ recursively as follows
\[
    \Delta(i,s) = - \sum_{j=i+1}^{i^*} \sum_{s' \geq s} \Delta(j,s') \cdot c(j,s',i,s).
\]
Furthermore, define
\[
    R_i := \sum_{j \geq i} \sum_{s \leq s^*} \Delta(j,s)
\]
and note that $R_i$ is the cumulative change to the objective function.

We can now prove by backward induction that:

\begin{align*}
%% first property
\text{(i)~ For $i = i^*$:}\quad
&\qquad\Delta(i^*,s^*) = - \delta \quad\text{for $s = s^*$, and}\\ &\qquad\Delta(i^*,s) = 0 \quad\text{for all $s \neq s^*$}\\[4pt]  
\text{\phantom{(i)~} For $i < i^*$:}\quad 
&\qquad\Delta(i,s) = %\textcolor{blue}{\delta \cdot \frac{1}{s^*} \left(\frac{s}{s^*}\right)^{i^*-2}} 
\delta \cdot \frac{1}{s^*+1} \prod_{t=1}^{i^*-2} \frac{s+t}{s^*+t+1}
\quad \text{for all $s \leq s^*$, and}\\[6pt]
&\qquad \Delta(i,s) = 0 \quad \text{for all $s > s^*$}
\end{align*}

Let's first do the base case. For $i = i^*$ the claim holds by definition. For $i = i^*-1$, we have 
\begin{align*}
\Delta(i^*-1,s) 
&= - \Delta(i^*,s^*) \cdot c(i^*,s^*,i^*-1,s)
= \delta \cdot \frac{1}{s^*+1} \cdot \prod_{t=1}^{i^*-2} \frac{s+t}{s^*+t+1},
\end{align*}
by the definition of $\Delta(i^*-1,s)$, $\Delta(i^*,s^*)$, and $c(i^*,s^*,i^*-1,s)$.

Let's do the inductive step. For this assume the claim holds for $i' > i$ and verify the inductive claim for $i$:
%\allowdisplaybreaks
\begin{align*}
    %% zeroth line
    &\Delta(i,s) \\
    %% first line
    \quad &= -  \sum_{j=i+1}^{i^*-1} \sum_{s' \geq s} \bigg( \Delta(j,s') \cdot c(j,s',i,s) \bigg) + \delta \cdot c(i^*,s^*,i,s) \\
    %% second line
    \quad &= - \sum_{j=i+1}^{i^*-1} \sum_{s' = s}^{s^*} \left( \delta \cdot \frac{1}{s^*+1} \cdot \prod_{t=1}^{i^*-2} \left(\frac{s'+t}{s^*+t+1}\right) \cdot \frac{1}{s'+1} \cdot \prod_{t=1}^{i-1} \left(\frac{s+t}{s'+t+1}\right)\right) + \delta \cdot \frac{1}{s^*+1} \cdot \prod_{t=1}^{i-1}\left( \frac{s+t}{s^*+t+1}\right)\\
    %% third line
    \quad &= -  \frac{\delta}{\prod_{t=1}^{i^*-1} (s^*+t)} \cdot \sum_{j=i+1}^{i^*-1} \left( \prod_{t=1}^{i-1} (s+t) \cdot  \sum_{s'=s}^{s^*}\left( \frac{\prod_{t=1}^{i^*-2}(s'+t)}{\prod_{t=1}^{i}(s'+t)}\right) \right)+ \delta \cdot \frac{1}{s^*+1} \cdot \prod_{t=1}^{i-1} \left(\frac{s+t}{s^*+t+1}\right)\\
    %% fourth line
    \quad &= - \frac{\delta}{\prod_{t=1}^{i^*-1} (s^*+t)} \cdot \sum_{j=i+1}^{i^*-1} \left( \prod_{t=1}^{i-1} (s+t) \cdot \frac{1}{i^*-i-1} \cdot \left( \frac{\prod_{t=1}^{i^*-1} (s^*+t)}{\prod_{t=1}^{i} (s^*+t)} -\frac{\prod_{t=1}^{i^*-2} (s+t)}{\prod_{t=1}^{i-1} (s+t)} \right)\right)\\ 
    %% fifth line
    \quad &\phantom{= - \sum_{j=i+1}^{i^*-1} \sum_{s' = s}^{s^*} \left( \delta \cdot \frac{1}{s^*+1} \cdot \prod_{t=1}^{i^*-2} \left(\frac{s'+t}{s^*+t+1}\right) \cdot \frac{1}{s'+1} \cdot \prod_{t=1}^{i-1} \left(\frac{s+t}{s'+t+1}\right)\right)}+ \delta \cdot \frac{1}{s^*+1} \cdot \prod_{t=1}^{i-1} \left(\frac{s+t}{s^*+t+1}\right)\\
    %% sixth line
    \quad &= - \frac{\delta}{\prod_{t=1}^{i^*-1} (s^*+t)} \cdot \prod_{t=1}^{i-1} (s+t) \cdot \left( \frac{\prod_{t=1}^{i^*-1} (s^*+t)}{\prod_{t=1}^{i} (s^*+t)} -\frac{\prod_{t=1}^{i^*-2} (s+t)}{\prod_{t=1}^{i-1} (s+t)} \right) + \delta \cdot \frac{1}{s^*+1} \cdot \prod_{t=1}^{i-1} \left( \frac{s+t}{s^*+t+1} \right)\\
    %% seventh line
    \quad &= -\delta \cdot \frac{1}{s^*+1} \cdot \prod_{t=1}^{i-1} \left(\frac{s+t}{s^*+t+1} \right)+ \delta \cdot \frac{1}{s^*+1} \cdot \prod_{t=1}^{i^*-2} \left(\frac{s+t}{s^*+t+1}\right) + \delta \cdot \frac{1}{s^*+1} \cdot \prod_{t=1}^{i-1} \left(\frac{s+t}{s^*+t+1}\right)\\
    %% eighth line
    \quad &= \delta \cdot \frac{1}{s^*+1} \cdot  \prod_{t=1}^{i^*-2} \left(\frac{s+t}{s^*+t+1}\right),\\
\end{align*}
where the second equation uses the inductive hypothesis and the fourth equality holds by Lemma \ref{lem:discrete_integral}.

Having established (i), we next show:
\begin{align*}
    %% second property
\text{(ii)}\quad \text{For all $i \leq i^*$:} 
&\qquad R_i = - \delta \cdot \frac{i-1}{i^*-1} \leq 0
\end{align*}

In fact, by the definition of $R_i$ and (i), we have
\begin{align*}
R_i 
= \left(\sum_{j = i}^{i^*-1} \sum_{s = 0}^{s^*} \Delta(j,s) \right) - \delta 
&= \delta \cdot \left( \sum_{j = i}^{i^*-1} \sum_{s=0}^{s^*}  \frac{1}{s^*+1} \prod_{t=1}^{i^*-2} \frac{s+t}{s^*+t+1} \right) - \delta \\
&= \delta \cdot \sum_{j=i}^{i^*-1} \left( \frac{
\sum_{s=0}^{s^*} \prod_{t=1}^{i^*-2} (s+t)
}{
\prod_{t=1}^{i^*-1} (s^*+t)
}\right) - \delta \\
&= \left(\frac{i^*-i}{i^*-1}-1\right) \delta = - \delta \cdot \frac{i-1}{i^*-1} \leq 0,
\end{align*}
where the last equality follows from Lemma~\ref{lem:discrete_integral}.

We now use (i) and (ii) to argue that the operation preserves feasibility and only improves the objective: 
We have chosen the recursion for the $\Delta(i,s)$ to satisfy the first constraint of the dual. From (i) we get that all the $\Delta(i,s)$ are non-negative, so we also satisfy the non-negativity constraints. From (ii) we get that $R_1 \leq 0$ so we only decrease the objective.

By repeatedly applying this operation we can conclude that the greedy solution is an optimal solution, just as we claimed.
\end{proof}

\section{Inductive Step in Proof of Theorem~\ref{thm:samples_explicit_dual}}\label{apx:inductive_step}

For the inductive step we will assume the claim is true for $s+1$. 

We want to show that:
\[
\sum_{s'=s}^{k} T(i,s') = \sum_{j=1}^{n-i} \left( {n-i \choose j} \cdot \frac{1}{j} \cdot  \prod_{\ell = 1}^{j} (k-s+\ell) \cdot \prod_{t=i}^{n-j-1} (s+t) \right).
\]

From the definition of $T(i,s')$ and the induction hypothesis we know that 
\begin{align*}
    \sum_{s' = s}^{k} T(i,s') &= T(i,s) + \sum_{s' = s+1} T(i,s')\\
    &= \left((n-i)\prod_{t=i+1}^{n-1} (s+t) - \sum_{j=1}^{n-i-1} \left((n-i) \cdot {n-i-1 \choose j} \cdot  \frac{1}{j(j+1)} \cdot \prod_{\ell = 1}^{j} (k-s+\ell) \cdot \prod_{t=i+1}^{n-j-1} (s+t)\right)\right)\\
    &\qquad\qquad + \left( \sum_{j=1}^{n-i} \left({n-i \choose j} \cdot  \frac{1}{j} \cdot \prod_{\ell = 1}^{j} (k-(s+1)+\ell) \cdot  \prod_{t=i}^{n-j-1} ((s+1)+t) \right) \right).
\end{align*}

We have:

\begin{align*}
%% first line
& \sum_{j=1}^{n-i} \left( {n-i \choose j} \cdot \frac{1}{j}  \cdot \prod_{\ell = 1}^{j} (k-s+\ell) \cdot   \prod_{t=i}^{n-j-1} (s+t) \right)\\
%% second line
%&\qquad\qquad= 
%\sum_{j=1}^{n-i-1} \left( {n-i \choose j} \cdot \frac{1}{j} \cdot \prod_{\ell = 1}^{j} (k-s+\ell) \cdot   \prod_{t=i}^{n-j-1} (s+t) \right) + \frac{1}{n-i} \cdot \prod_{\ell=1}^{n-i}(k-s+\ell)\\
%% third line
&\qquad\qquad= 
\sum_{j=1}^{n-i-1} \left( {n-i \choose j} \cdot \frac{1}{j}  \cdot(s+i)\cdot \prod_{\ell = 1}^{j} (k-s+\ell) \cdot   \prod_{t=i+1}^{n-j-1} (s+t) \right) + \frac{1}{n-i} \cdot \prod_{\ell=1}^{n-i}(k-s+\ell)\\
%% fourth line
&\qquad\qquad= 
\sum_{j=1}^{n-i-1} \left( {n-i \choose j} \cdot \frac{1}{j}  \cdot(s+n-j-n+j+i)\cdot \prod_{\ell = 1}^{j} (k-s+\ell) \cdot   \prod_{t=i+1}^{n-j-1} (s+t) \right) + \frac{1}{n-i}\prod_{\ell=1}^{n-i}(k-s+\ell)\\
%% fifth line
&\qquad\qquad= (k-s+1)\cdot (n-i) \cdot \prod_{t=i+1}^{n-1} (s+t) + \sum_{j=2}^{n-i-1} \left( {n-i \choose j} \cdot \frac{1}{j}  \cdot(s+n-j)\cdot \prod_{\ell = 1}^{j} (k-s+\ell) \cdot   \prod_{t=i+1}^{n-j-1} (s+t) \right)   \\
&\qquad\qquad\qquad-\sum_{j=1}^{n-i-1} \left( {n-i \choose j} \cdot \frac{1}{j}  \cdot(n-j-i)\cdot \prod_{\ell = 1}^{j} (k-s+\ell) \cdot   \prod_{t=i+1}^{n-j-1} (s+t) \right) + \frac{1}{n-i} \cdot \prod_{\ell=1}^{n-i}(k-s+\ell)\\
%% sixth line
%&\qquad\qquad= (k-s+1)\cdot (n-i) \cdot \prod_{t=i+1}^{n-1} (s+t) + \sum_{j=1}^{n-i-2} \left( {n-i \choose j+1} \cdot \frac{1}{j+1}  \cdot(s+n-j-1)\cdot \prod_{\ell = 1}^{j+1} (k-s+\ell) \cdot   \prod_{t=i+1}^{n-j-2} (s+t) \right)  \\
%&\qquad\qquad\qquad-\sum_{j=1}^{n-i-1} \left( {n-i \choose j} \cdot \frac{1}{j}  \cdot(n-j-i)\cdot \prod_{\ell = 1}^{j} (k-s+\ell) \cdot   \prod_{t=i+1}^{n-j-1} (s+t) \right) + \frac{1}{n-i}\prod_{\ell=1}^{n-i}(k-s+\ell)\\
%% seventh line
&\qquad\qquad= (k-s+1)\cdot (n-i)\cdot \prod_{t=i+1}^{n-1} (s+t) +
\sum_{j=1}^{n-i-2} \left( {n-i \choose j+1} \cdot \frac{1}{j+1}  \cdot \prod_{\ell = 1}^{j+1} (k-s+\ell) \cdot   \prod_{t=i+1}^{n-j-1} (s+t) \right)  \\
&\qquad\qquad\qquad-\sum_{j=1}^{n-i-1} \left( {n-i \choose j} \cdot \frac{1}{j}  \cdot(n-j-i)\cdot \prod_{\ell = 1}^{j} (k-s+\ell) \cdot   \prod_{t=i+1}^{n-j-1} (s+t) \right) + \frac{1}{n-i}\prod_{\ell=1}^{n-i}(k-s+\ell)\\
%% eighth line
&\qquad\qquad= (k-s+1) \cdot (n-i) \cdot \prod_{t=i+1}^{n-1} (s+t) +
\sum_{j=1}^{n-i-1} \left( {n-i \choose j+1} \cdot \frac{1}{j+1}  \cdot \prod_{\ell = 1}^{j+1} (k-s+\ell) \cdot   \prod_{t=i+1}^{n-j-1} (s+t) \right)   \\
&\qquad\qquad\qquad-\sum_{j=1}^{n-i-1} \left( {n-i \choose j} \cdot \frac{1}{j}  \cdot(n-j-i)\cdot \prod_{\ell = 1}^{j} (k-s+\ell) \cdot   \prod_{t=i+1}^{n-j-1} (s+t) \right) \\
%% ninth line
&\qquad\qquad= (k-s+1) \cdot (n-i) \cdot \prod_{t=i+1}^{n-1} (s+t) +
\sum_{j=1}^{n-i-1} \left( {n-i \choose j+1} \cdot \frac{1}{j+1} \cdot (k-s+j+1) \cdot \prod_{\ell = 1}^{j} (k-s+\ell) \cdot   \prod_{t=i+1}^{n-j-1} (s+t) \right)  \\
&\qquad\qquad\qquad-\sum_{j=1}^{n-i-1} \left( {n-i \choose j} \cdot \frac{1}{j}  \cdot(n-j-i)\cdot \prod_{\ell = 1}^{j} (k-s+\ell) \cdot   \prod_{t=i+1}^{n-j-1} (s+t) \right) \\
%% tenth line
&\qquad\qquad= (k-s+1) \cdot (n-i) \cdot \prod_{t=i+1}^{n-1} (s+t) +
\sum_{j=1}^{n-i-1} \left( {n-i \choose j+1} \cdot \frac{1}{j+1} \cdot (k-s) \cdot \prod_{\ell = 1}^{j} (k-s+\ell) \cdot   \prod_{t=i+1}^{n-j-1} (s+t) \right) +\\
&\qquad\qquad\qquad+\sum_{j=1}^{n-i-1} \left( {n-i \choose j+1} \cdot \prod_{\ell = 1}^{j} (k-s+\ell) \cdot   \prod_{t=i+1}^{n-j-1} (s+t) \right)  \\
&\qquad\qquad\qquad-\sum_{j=1}^{n-i-1} \left( {n-i \choose j} \cdot  \frac{1}{j}  \cdot(n-j-i)\cdot \prod_{\ell = 1}^{j} (k-s+\ell) \cdot   \prod_{t=i+1}^{n-j-1} (s+t) \right) \\
% eleventh line
&\qquad\qquad= (k-s) \cdot (n-i) \cdot \prod_{t=i+1}^{n-1} (s+t) +
\sum_{j=1}^{n-i-1} \left( {n-i \choose j+1} \cdot \frac{1}{j+1} \cdot (k-s) \cdot \prod_{\ell = 1}^{j} (k-s+\ell) \cdot   \prod_{t=i+1}^{n-j-1} (s+t) \right) +\\
&\qquad\qquad\qquad+(n-i) \cdot \prod_{t=i+1}^{n-1} (s+t) +\sum_{j=1}^{n-i-1} \left( {n-i \choose j+1} \cdot  \prod_{\ell = 1}^{j} (k-s+\ell) \cdot   \prod_{t=i+1}^{n-j-1} (s+t) \right)  \\
&\qquad\qquad\qquad-\sum_{j=1}^{n-i-1} \left( {n-i \choose j} \frac{1}{j}  \cdot(n-j-i)\cdot \prod_{\ell = 1}^{j} (k-s+\ell) \cdot   \prod_{t=i+1}^{n-j-1} (s+t) \right) \\
%% UP TO HERE
%% twelfth line
&\qquad\qquad= (k-s) \cdot (n-i) \cdot \prod_{t=i+1}^{n-1} (s+t) +
\sum_{j=1}^{n-i-1} \left( {n-i \choose j+1} \cdot  \frac{1}{j+1} \cdot (k-s) \cdot \prod_{\ell = 1}^{j} (k-s+\ell) \cdot   \prod_{t=i+1}^{n-j-1} (s+t) \right) +\\
&\qquad\qquad\qquad+(n-i) \cdot \prod_{t=i+1}^{n-1} (s+t) -
\sum_{j=1}^{n-i-1} \left( (n-i) {n-i-1 \choose j} \frac{1}{j(j+1)} \cdot \prod_{\ell = 1}^{j} (k-s+\ell) \cdot   \prod_{t=i+1}^{n-j-1} (s+t) \right).
\end{align*}

We furthermore have that
\begin{align*}
    %% first line
    &(k-s) \cdot (n-i) \cdot \prod_{t=i+1}^{n-1} (s+t) +
    \sum_{j=1}^{n-i-1} \left( {n-i \choose j+1} \cdot  \frac{1}{j+1} \cdot (k-s) \cdot \prod_{\ell = 1}^{j} (k-s+\ell) \cdot   \prod_{t=i+1}^{n-j-1} (s+t) \right)\\
    %% second line
    &\qquad\qquad= 
    (k-s) \cdot (n-i) \cdot \prod_{t=i+1}^{n-1} (s+t) +
    \sum_{j=2}^{n-i} \left( {n-i \choose j} \cdot  \frac{1}{j} \cdot \prod_{\ell = 0}^{j-1} (k-s+\ell) \cdot   \prod_{t=i+1}^{n-j} (s+t) \right)\\
    %% third line
    &\qquad\qquad= 
    (k-s) \cdot (n-i) \cdot \prod_{t=i+1}^{n-1} (s+t) +
    \sum_{j=2}^{n-i} \left( {n-i \choose j} \cdot  \frac{1}{j} \cdot \prod_{\ell = 1}^{j} (k-(s+1)+\ell) \cdot   \prod_{t=i}^{n-j-1} ((s+1)+t) \right)\\
    %% third line
    &\qquad\qquad= 
    \sum_{j=1}^{n-i} \left( {n-i \choose j} \cdot  \frac{1}{j} \cdot \prod_{\ell = 1}^{j} (k-(s+1)+\ell) \cdot   \prod_{t=i}^{n-j-1} ((s+1)+t) \right),
\end{align*}
which completes the proof.

\section{Full Proof of Theorem~\ref{thm:binary_thresholds}}\label{apx:binary}

\begin{proof}[Full Proof of Theorem~\ref{thm:binary_thresholds}]
From Proposition~\ref{prop:greedy_thresholds} we know that it suffices to show that the greedy algorithm finds an optimal solution $u_i^s$ to the dual LP, and that the $u_i^s$ are non-decreasing in $i$.

For ease of reference, let's recall the equation that defines the greedy algorithm (equation (\ref{eq:dual_dp})). For $i = n$ and all $s$ the greedy algorithm sets $u_n^s = a(n,s)$ and for $i < n$ and all $s$ it sets
\begin{equation}\label{eq:dual_dp_yes_no}
u_i^s = \max\bigg(0, a(i,s) -  \sum_{j>i} \sum_{s'}  u_j^{s'} \cdot  \c(i,s)
\bigg).
\end{equation}
where $a(i,Y) = p/n$, $a(i,N) = (1-p)/N$, $\c(i,Y) = (1-p')/i$, and $\c(i,N) = p'/i$.
%Moreover, $c(j,s',i,s)$ is actually independent of $j$ and $s'$ and equal to:
%$$c(i,s,j,Y) = \frac{1-p'}{j} \quad \text{and} \quad  c(i,s,j,N) = \frac{p'}{j}$$

Monotonicity of the $u_i^s$ constructed via equation (\ref{eq:dual_dp_yes_no}) follows from the facts that $a(i,s)$ does not depend on $i$, and that $\hat{c}(i,s)$ is decreasing in $i$.

For optimality assume that we have a dual solution that is not of the form above. Take the $(i^*, s^*)$ point with largest possible $i^*$ such that  equation \eqref{eq:dual_dp_yes_no} is not an equality. We will show how to change the solution to make it an equality without hurting feasibility or increasing the dual objective. 

Specifically, consider changing $u_i^s$ to $u_i^s + \Delta(i,s)$ where $\Delta(i^*, s^*) = -\delta$ for $$\delta := u(i^*, s^*) - \max\bigg(0, a(i^*, s^*) -  \sum_{j>i^*} \sum_{s'}  u_j^{s'} \cdot \c(i^*, s^*) \bigg)$$ and $\Delta(i,s) = 0$ for all other $i \geq  i^*$. For $i < i^*$ we can define $\Delta(i,s)$ recursively as follows:
$$\Delta(i,s) = -\sum_{j=i+1}^{i^*} \sum_{s'} \Delta(j,s') \cdot  \c(i,s).$$
We will show (by a backwards recursion) that for each $i < i^*$ we have:
$$\text{(i) } \Delta(i,s) \geq 0 \qquad \text{and} \qquad \text{(ii) } R_i := \sum_{j \geq i} \sum_s \Delta(j,s) \leq 0$$
Let's first check that base case $i=i^*-1$. For this case, we have:
$$\Delta(i^*-1,s) = - \c(i^*-1,s) \cdot \Delta(i^*,s^*) = \c(i^*-1,s) \cdot \delta \geq 0$$
which shows (i). For (ii), note that:
$$R_{i^*-1} = -\delta + \sum_{s} \c(i^*-1,s) \cdot \delta.$$
Since $\c(i,s)$ represent probability distributions, we have that  $\sum_s \c(i,s) \leq 1$. This directly implies that $R_{i^*-1} \leq 0$.\\

For the induction step, note that $$\Delta(i,s) = -\c(i,s) \cdot R_{i+1}.$$
Since $R_{i+1} \leq 0$ then $\Delta(i,s) \geq 0$ showing (i). For (ii) observe that: 
$$R_i = \sum_s \Delta(i,s) + R_{i+1} = -R_{i+1} \cdot \sum_s \c(i,s) + R_{i+1}  = R_{i+1} (1-\sum_s \c(i,s)) \leq 0$$
since $\c(i,s)$ is a probability distribution.

Now that we established (i) and (ii) observe that $(ii)$ for $i=1$ implies that the objective function doesn't increase, so if the original solution was optimal, the transformed solution will also be optimal. By (i) we establish that $u_i^s + \Delta(i,s)$ is still non-negative. Finally, note that $\Delta(i,s)$ is defined precisely to make the constraints in the dual LP feasible. Hence, by applying this transformation repeatedly starting form any optimal solution to the dual, we arrive at an optimal solution of the dual that is obtained by the greedy algorithm in equation \eqref{eq:dual_dp_yes_no}.
\end{proof}

\end{document}